\documentclass[final,5p,times,twocolumn]{elsarticle}

\usepackage{amssymb}
\usepackage{amsmath}
\usepackage{amsthm}
\usepackage{nicefrac, xfrac}
\usepackage{xcolor}
\usepackage{multirow}
 \usepackage{url}
 \usepackage{fancyhdr}

\newcommand{\tilda}{~}
\newcommand{\R}{\mathbb{R}}
\newcommand{\RR}{\mathbb{R}^{3\times3}}

\newcommand{\gr}{\}}
\newcommand{\gl}{\{}

\newtheorem{proposition}{Proposition}
\newtheorem{remark}{Remark}
\newtheorem{corollary}{Corollary}
\newtheorem{definition}{Definition}
\newtheorem{lemma}{Lemma}

\journal{Control Engineering Practice}

\begin{document}

\begin{frontmatter}



\title{Artificial Potential Field and Sliding Mode Control for Spacecraft Attitude Maneuver with Actuation and Pointing Constraints}


\author[label1]{Mauro Mancini\corref{cor1}}\ead{mauro.mancini@polito.it}

\author[label1]{Dario Ruggiero}\ead{dario.ruggiero@polito.it} 

\affiliation[label1]{organization={Department
of Mechanical and Aerospace Engineering, Politecnico di Torino},
            addressline={Corso Duca degli Abruzzi 24}, 
            city={Torino},
            postcode={10129}, 
            country={Italy}}
            
\cortext[cor1]{Corresponding author}
\fntext[]{Received 31 July 2024, Revised 16 April 2025, Accepted 16 April 2025
\newline
Digital Object Identifier (DOI): \url{10.1016/j.conengprac.2025.106373}}

\begin{abstract}
This study investigates the combination of guidance and control strategies for rigid spacecraft attitude reorientation, while dealing with forbidden pointing constraints, actuators limitations and system uncertainties. Reasons are related to the presence of bright objects in space that may damage sensitive payloads installed on the spacecraft, while saturations of attitude actuators may compromise the closed-loop system stability. In addition, the spacecraft attitude dynamics is typically affected by parametric uncertainties, external disturbances, and system nonlinearities, which cannot be neglected in the analysis. 
In this article, the problem of spacecraft reorientation under pointing and actuation constraints is solved with a strategy combining Artificial Potential Field (APF) and Sliding Mode Control (SMC). 
Following rigorous Lyapunov analysis, closed-form expressions for APF/SMC gains are obtained, i.e. explicit mathematical expressions that directly provide the control gain values without the need for iterative or recursive calculations, while accounting for angular velocity and control torque limitations, external disturbances, and inertia uncertainties.
The robustness of the proposed control strategy against inertia uncertainties, external disturbances, and actuator constraints is validated through Monte Carlo simulations in a high-fidelity attitude dynamics simulator, while $\mu$-analysis is used to assess local stability properties and quantify the system’s robustness margins.
These results demonstrate the practical feasibility of the proposed control method in real-world scenarios, highlighting its robustness in complex, uncertain environments typical of space operations.
\end{abstract}



\end{frontmatter}

\textcopyright 2025. This manuscript version is made available under the CC-BY-NC-ND 4.0 license \url{http://creativecommons.org/licenses/by-nc-nd/4.0/}



\section{Introduction}
\label{sec:intro}
The importance of the spacecraft attitude control problem has prompted the scientific community to delve into this field of research since the dawn of space missions. Indeed, the literature proposes a variety of non-linear control techniques addressing unconstrained attitude-maneuver control \cite{609665,4554017,6034642,7828857}. 
In contrast, attitude-maneuver control involving pointing constraints has only been examined in a few recent research works. Pointing constraints arise from the need to shield sensitive payloads from bright sources of energy, thus leading to the definition of forbidden zones in 3-dimensional space \cite{1298998}. Then, the attitude controller must ensure that the instrument does not enter the forbidden zones while maneuvering the spacecraft to the desired orientation. 

Solutions for spacecraft reorientation with forbidden zones can be divided into path planning methods, trajectory optimization methods, and Artificial Potential Field (APF) methods. A comprehensive review of these methods and their applications can be found in \cite{8890837, HUA2024108738}, and the references therein. The first two approaches can guarantee high performance, but are unsuitable for real-time onboard implementation because they require high computational cost. From this point of view, APF is a viable alternative due to its computational ease. APF methods shape the environment with artificial potentials by assigning a global attractive minimum at the target reference and high values around the forbidden zones. Then, the potential function is incorporated in the attitude controller design to ensure the slew trajectories of the spacecraft runs along the gradient of the artificial potentials, thus converging to the desired state while avoiding the forbidden zones. 

In the literature, a variety of potential functions have been proposed to realise attitude maneuvers with pointing constraints, such as Gaussian functions \cite{Mclnnes}, logarithmic barrier functions \cite{6978863}, and exponential functions \cite{Avanzini}. 
Furthermore, many control schemes are proposed to complement the APF methods, such as PD control \cite{doi:10.2514/1.G003606}, backstepping control \cite{6978863}, and Sliding Mode Control (SMC) \cite{SHEN2018157}. 
A lot of research has been done over the years to improve the preliminary works on APF-based attitude planning algorithms. The research efforts resulted in new methods for solving earlier problems related to  multiple pointing constraint zones \cite{6978863}, local minima and unwinding phenomenon \cite{doi:10.2514/1.G003606}, angular rate \cite{SHEN2018157} and  {control torques limits \cite{Hu}}, and time delays \cite{XU2022107885}. 

 In the works cited above, the overall potential is generated by combining attractive potential towards desired attitude and repulsive potential from forbidden pointing direction. Furthermore, the overall potential is scaled through proper gains, which shape the size of forbidden zones and the magnitude of attraction force. However, none of the previous studies provides analytical formulations for the regulation of the gains of the attractive and repulsive fields, which is crucial for achieving the desired balance between repulsive and attractive actions. 
Furthermore, most of the works in the literature do not address the issue of actuator constraints in spacecraft reorientation with forbidden zones, which means that the results obtained may not be replicable in a real-world system. 
In \cite{8854817}, a barrier Lyapunov function is used to constrain the satellite's angular velocity during the maneuver, but no limits are imposed on the control torque. 
In \cite{Hu}, a numerical method is presented to ensure the boundedness of the control torque, but the problem of limiting the maneuvering angular velocity is not addressed. This could lead to saturation of low-rate gyro actuators and, consequently, system instability. 
Similarly, in \cite{XU2022107885}, the control torque is bounded by solving two optimization problems, but no limit is imposed on the angular velocity. 
An analytical strategy for limiting angular velocity is proposed in \cite{TIAN2022107332}, but does not include control torque limitation. 
A strategy to simultaneously satisfy the limits on control torque and angular velocity is proposed in \cite{9440030}. However, the control gains are obtained through a numerical method rather than in closed-form expressions, which allow for easier implementation and generalization.

In the context of autonomous attitude maneuvers for satellites, this work proposes a method to obtain robust stabilizing control gains {through an explicit formulation} to achieve maneuvers that comply with pointing and actuator constraints, even in the presence of disturbances and uncertainties.
In particular, this  {paper} proposes APF/SMC guidance and control strategies to realize rest-to-rest spacecraft attitude maneuvers complying with both pointing and actuation constraints.
 {In the literature, some recent works proposed strategies based on the combination of APF/SMC to address the problem of spacecraft reorientation under pointing constraints. However, none of these works provide closed-form expressions to obtain control gains that stabilise the system while respecting the actuation constraints. 
In \cite{10109162, GUAN202350}, the APF is introduced into the design of the sliding surface to complete anti-unwinding reorientation with multiple attitude constraints, but these strategies do not account for control torque or angular velocity limitations. 
APF and SMC are combined in a similar way in \cite{SHEN2018157} and \cite{doi:10.2514/1.G005026}, but control torques saturation are not considered in the first work, while the second one does not consider angular velocity limitation. 
An APF/SMC strategy that simultaneously satisfies angular velocity and control torque limits is presented in \cite{10354510}. However, in this work the control gains are obtained by numerical methods.}

Compared to the works in the literature, closed-form expressions are obtained in this paper for the selection of control gains of an APF/SMC based guidance and control strategy for spacecraft reorientation under pointing constraints. Furthermore, the analytical equations derived in this work allow for the simultaneous satisfaction of angular velocity and control torque limits, even in the presence of inertia uncertainties and external disturbances. 
In particular, the APF gains are selected so that the satellite's angular velocity does not overload the actuators angular momentum, while the SMC gains are derived to ensure that the control input avoids control torque saturation.  
Furthermore, analytical equations are derived to adjust the repulsive gains in relation to the width of the forbidden zone. These relations allow the repulsive field to be precisely adjusted so that the repulsive action prevails above the attractive one at a selectable angular distance from the forbidden direction. 
In the proposed strategy, the attractive layer of the APF is built with the quaternion describing the shortest rotation from actual to desired orientation. Instead, the repulsive potential is based on the quaternions describing the shortest rotation that would bring the boresight vector onto the forbidden direction axis. 
Obviously, both sets of quaternions above are calculated employing the forbidden directions and the direction of the boresight vector. 
Furthermore, the proposed APF strategy avoids the unwinding phenomenon and guarantees that the shorter-path eigenaxis rotation maneuver is performed when there are no forbidden zones affecting the calculation of the trajectory. 
The APF strategy is combined with SMC, which is a nonlinear control technique with remarkable properties of precision and robustness \cite{utkin2013sliding}. The SMC design process consists in two steps. First, a sliding surface (a sub-space of the system's states) is selected with attractive convergence properties. Then, a discontinuous control law is designed to confine the system’s trajectories on the sliding surface. This work employs boundary layer SMC \cite{slotine1983tracking} to avoid high-frequency oscillations in the command line (chattering) and streamlines a strategy for tuning the gains of the combined APF/SMC approach to ensure compliance with actuation constraints.  
It is worth noting that while the adoption of a boundary layer SMC may slightly weaken the ideal convergence properties and robustness—since the sliding variable is confined within a thin boundary layer rather than driven exactly to zero—our careful tuning ensures that the norm of the vector part of the quaternion error remains bounded. Moreover, the boundary layer approach helps prevent excessive stress on the actuators, which is particularly important given the system's physical limitations.
Furthermore, although higher-order SMC techniques (e.g., the Super-Twisting algorithm) are known to reduce chattering through continuous control actions, they can be particularly sensitive to low sampling frequencies when implemented on digital computers \cite{UTKIN202010244, Boiko2009, shtessel2014sliding}. Given the limited sampling rate imposed by current onboard spacecraft hardware, first-order SMC with a boundary layer provides a more reliable and practical solution for this application. 
Another interesting aspect of this  {paper} is the derivation of angular velocity and control torque constraints based on real actuators, i.e. a pyramid cluster of Reaction Wheels (RWs). This aspect allows to validate the integration of the proposed APF/SMC strategy with a real system.
Then, the guidance and control strategy is validated by numerical simulations performed with a three degree-of-freedom (DOF) orbital simulator. The latter includes the forbidden zones, the non-linear equations of attitude dynamics and kinematics, the APF/SMC algorithms, and the model of RWs cluster. Furthermore, hardware and software constraints are taken into account in terms of variation of the update (switching) frequency of the closed-loop system with the combined algorithms. 
Robust stability verification is an important part of space Guidance Navigation and Control (GNC) system design \cite{kron2010mu,rodrigues2022modeling,chan2017verifying}. Some recent approaches, such as Integral Quadratic Constraint \cite{megretski1997system}, allow to perform robust stability verification of nonlinear systems. However,  generally these methods require higher computational effort in comparison with classical approaches \cite{chaudenson2013robustness}. In \cite{pagone2020gnc}, a novel practical approach to address nonlinear system robust stability verification is discussed using $\mu$-analysis \cite{young1991mu}. It is based on the classical linear fractional transformation of the system, while system nonlinearities are accounted by considering input-output relationship with the linearized plant. In this work, $\mu$-analysis is carried out to analyze the robust stability of the proposed closed-loop system, comparing the results for the linearized model and the nonlinear model following the approach discussed in \cite{pagone2020gnc}. {In this context, the $\mu$-analysis is used as a practical approach to assess the system’s robustness limits, effectively defining a performance index that quantifies how much uncertainty the closed-loop system can tolerate before becoming unstable.} A simulation campaign is conducted as part of the robust stability verification, proving the reliability of the results of $\mu$-analysis, and evaluating algorithm performance and effectiveness in compliance with system uncertainties.

In reference to autonomous spacecraft constrained attitude maneuvers domain, the main contributions of this paper are given by the combination of both theoretical and numerical analysis as follow: 
\begin{enumerate}
    \item Derivation of control gain expressions: Closed-form expressions are derived for the selection of stabilizing control gains in the APF/SMC strategy, accounting for key factors such as the size of the forbidden zones, the magnitude of inertia uncertainties, external disturbances, and actuator constraints. 
    \item Assessment of sensitivity to uncertainties: The sensitivity of the system’s local closed-loop stability to specific uncertainties and partial nonlinearities is evaluated, giving insights into its robustness under practical operative conditions.
\end{enumerate}
Additionally, the proposed APF/SMC strategy is integrated within a realistic system setup to assess implementation feasibility, supported by a Lyapunov stability analysis that accounts for actuation constraints, parametric uncertainties, and system nonlinearities.
The effectiveness of the guidance and control strategy is validated in a high-fidelity simulation environment, based on a real satellite model (DEMETER from CNES), which includes hardware constraints and a detailed actuation system model with saturation limits and filtering effects.
Finally, the robustness of the proposed control strategy against system nonlinearities and parameter uncertainties is {evaluated by $\mu$-analysis, and} verified through Monte Carlo simulations, demonstrating the strategy's effectiveness across a wide range of scenarios.

The article is organized as follows.  Section II gives the notation used throughout the article and specify the definitions and lemmas for system stability. Section III describes the  spacecraft attitude dynamics, the cluster of RWs, and the pointing constraints. The guidance and control strategy is in Section IV, where the proof of closed-loop stability and the robust stability verification are also detailed. The simulation scenario and the results are in Section V. Finally, some concluding remarks are in Section VI.

\begin{table}[]
    \centering
    \caption{Nomenclature}
    \begin{tabular}{|c|l|}
    \hline
    \textbf{Symbol} & \textbf{Description}\\
    \hline
      $q,\omega$  & Real quaternions and angular speed. \\
    \hline
     $q_d$ & Desired quaternions.\\
     \hline
     \multirow{2}{*}{$\hat{m}_I,\hat{n}_j$} & Orientation of sensor axis and of $j-th$ \\
     & forbidden axis in the inertial frame.\\
     \hline
     \multirow{2}{*}{$\tilde{q}_j$} & Orientation of the sensor axis with respect\\
     &  to the $j$-th forbidden axis.\\
     \hline
     \multirow{3}{*}{$\omega_a,\omega_{r,j},\omega^*,q^*$} & APF-based guidance outputs: attractive  \\
     & and repulsive layers, reference angular \\
     & speed and quaternions.\\
     \hline
     \multirow{2}{*}{$e_\epsilon,e_\omega$} & Quaternions and angular velocity errors\\
     & with respect to reference $q^*,\omega^*$.\\
     \hline
     \multirow{3}{*}{$u,\tau,h$} & Control input, torque generated by the\\
     & reaction wheels, and total angular\\
     &momentum stored into the wheels.\\
     \hline
     \multirow{2}{*}{$\tau^w,h^w$} & Torque and angular momentum of each\\
     &reaction wheel.\\
     \hline
     $I^*,\Delta_I$ & Inertia tensor and its uncertainties.\\
     \hline
     $d$ & External disturbances.\\
     \hline
    \end{tabular}
    \label{tab:my_label}
\end{table}

\section{ {PRELIMINARIES}}
This section presents first the notation used throughout the article.  {Then, it provides the the definitions of stability and the lemmas that will be used subsequently to prove the stability of the closed-loop system.}

\subsection{Notation}
    The following notations are used throughout the article.  {Let $x=(x_1,~\dots,~x_n)^T\in\R^n$}, then $\|x\|$ and  $\|x\|_1$ indicate the $l^2-$ vector norm and the $l^1-$vector norm of $x$, respectively, with $\|x\|\leq\|x\|_1$. Consequently, $\|A\|$, $\|A\|_1$ are the $l^2-$ induced matrix norm and the $l^1-$induced matrix norm of  {$A\in\R^{n\times n}$}, respectively. Given a further matrix  {$B\in\R^{n\times n}$}, the notation $A>B$ indicates that each element $a_{ij}$ and $b_{ij}$ of $A$ and $B$, respectively, satisfy $a_{ij}>b_{ij}$. On the contrary,  $A<B$ indicates that $a_{ij}<b_{ij}$ for each $i,j$. Then,  { $\left|x\right|=\left(\left|x_1\right|,~\dots,~\left|x_n\right|\right)^T\in\R^n$}  indicates the element wise absolute value of $x$. Consistently, given a scalar $c\in\R$, $\left|c\right|$ indicates its absolute value. Instead, $\Hat{x}=x/\|x\|$ indicates the normalized unit vector, i.e. $\|\Hat{x}\|=1$.  {Then, $\mathbb{I}_n$ identifies the identity matrix of size $n\times n$, while $\mathbf{1}_n=(1,~\dots,~1)^T\in\R^n$ and $\mathbf{0}_n=(0,~\dots,~0)^T\in\R^n$ are vectors of dimension $n$ whose elements are all $1$ and $0$, respectively. Given $c\in\R$ and $\mathbf{1}_n\in\R^n$ as defined above,  $diag\left(c\mathbf{1}_n\right)$ indicates an n-by-n diagonal matrix with the vector $c\mathbf{1}_n$ on the diagonal.}

Consider now  {$x\in\R^3$.} Then, $x^\times\in\RR$ denotes the skew symmetric matrix of $x$: $x^\times=[0,~-x_3,~x_2;x_3,~0,~-x_1;-x_2,~x_1,~0]$. Furthermore, let  {$\Hat{x}\in\R^3$} and $\Hat{y}\in\R^3$ be two unit vectors. Then, $\delta=ang(\Hat{x},\Hat{y})$ indicates the angle between $\Hat{x}$ and $\Hat{y}$ such that $\Hat{x}^T \Hat{y}=cos(\delta)$. In addition, $q=[\eta,~\epsilon^T]^T\in\R^4$ indicates an attitude quaternion $(\|q\|=1)$, where $\eta\in\R$ and $\epsilon\in\R^3$ are the scalar and the vectorial part of $q$, respectively. The quaternion conjugate of $q$ is denoted with $\overline{q}=[\eta,~-\epsilon^T]^T$ and $\otimes$ indicates the quaternion multiplication. Finally, $sign_+(\cdot)$ assumes the value -1 for negative input element and +1 otherwise. Frequently used symbols are reported in Table \ref{tab:my_label}.

\subsection{ {Definitions and Lemmas}}
     {Consider a dynamic system modeled as
    \begin{equation}
        \Dot{x}(t)=f(x(t)),\quad x(t)\in\mathcal{X}\subset\R^n,
        \label{eq:AutoSys}
    \end{equation}
    where $f$ is assumed to be locally Lypschitz and $\phi(t,x_0)$ is used to denote the solution of (\ref{eq:AutoSys}) at time $t$ with initial condition $x(0)=x_0$. 
    Consider now a set $\mathcal{G}\subset\mathcal{X}$ as a ball of radius $\rho>0$ centered in the origin $x=0$, i.e.
    \begin{equation}
        \mathcal{G}=\{x\in\mathcal{X}:\tilda\|x\|<\rho\},
        \label{eq:Gamma_for_Stability}
    \end{equation}
    a scalar $\varepsilon>0$, $y\in\mathcal{G}$, and define:
    \begin{itemize}
        \item The point-to-set distance of $x$ to $\mathcal{G}$ as $\|x\|_\mathcal{G}:=\text{inf}\{\|x-y\|\}$.
        \item The set $B_{\varepsilon}(\mathcal{G}):=\{ x\in \mathcal{X}:~\|x\|_\mathcal{G}<\varepsilon \}$.
        \item The set $\phi(\R_+,\mathcal{G}):=\{\phi(t,x_0):\tilda t\in\R_+,\tilda x_0\in \mathcal{G}\}$.
    \end{itemize}   
    Now, let $\mathcal{G}\subset\mathcal{X}$ be a closed positively invariant set for \eqref{eq:AutoSys}, $\mathcal{N}(\mathcal{G})$ an open neighborhood of $\mathcal{G}$, and consider the following definitions:
    \begin{definition} (Stability of compact sets)\\
        \begin{enumerate}[(i)]
            \item $\mathcal{G}$ is stable for \eqref{eq:AutoSys} if for all $\varepsilon>0$ there exists a neighborhood $\mathcal{N}(\mathcal{G})$ such that $\phi(\R_+,\mathcal{N}(\mathcal{G}))\subset B_\varepsilon(\mathcal{G})$.
            \item $\mathcal{G}$ is attractive for \eqref{eq:AutoSys} if there exists a neighborhood $\mathcal{N}(\mathcal{G})$ such that $\text{lim}_{\tilda t\to+\infty}\|\phi(t,x_0)\|_\mathcal{G}=0$ for all $x_0\in\mathcal{N}(\mathcal{G})$. 
            \item $\mathcal{G}$ is globally attractive for \eqref{eq:AutoSys} if is attractive for $\mathcal{N}(\mathcal{G})=\mathcal{X}$.
            \item $\mathcal{G}$ is (globally) asymptotically stable for \eqref{eq:AutoSys} if is stable and (globally) attractive for \eqref{eq:AutoSys}.
            \item $\mathcal{G}$ is  finite-time attractive for \eqref{eq:AutoSys} if there exist a neighborhood $\mathcal{N}(\mathcal{G})$ and 
            a function $t_r(x_0)\in[0,\infty)$, called the settling-time function, such that  $\text{lim}_{\tilda t\to t_r}\|\phi(t,x_0)\|_\mathcal{G}=0$ for all $x_0\in\mathcal{N}(\mathcal{G})$.
            \item $\mathcal{G}$ is globally  finite-time attractive for \eqref{eq:AutoSys} if is  finite-time attractive for $\mathcal{N}(\mathcal{G})=\mathcal{X}$.
            \item $\mathcal{G}$ is (globally) finite-time stable for \eqref{eq:AutoSys}, with settling time $t_r(x_0)=\text{inf}\{t\in\R_+:\tilda\|\phi(t,x_0)\|_\mathcal{G}=0\}$, if is stable and (globally) finite-time attractive for \eqref{eq:AutoSys}.
        \end{enumerate}\qed
    \end{definition}
    Then, the following Lemmas can be stated:
    \begin{lemma} [Theorem 4.2 in \cite{khalil2002nonlinear}]
            If there exists a continuously differentiable function $V:~\mathcal{N}(\mathcal{G})\to\R$ such that: 
            \begin{align}
                &V(0)=0\text{ and }V(x)>0\tilda,\forall \tilda x\in\mathcal{N}(\mathcal{G}){\setminus}0,\label{eq:V_AS}
                \intertext{and the following holds along the trajectories of \eqref{eq:AutoSys}:}
                &\dot{V}(x)<0,~\forall x\in\mathcal{N}(\mathcal{G})\setminus\mathcal{G},\label{eq:dotV_AS}
            \end{align}
            then the set $\mathcal{G}$ in \eqref{eq:Gamma_for_Stability} is asymptotically stable for the system (\ref{eq:AutoSys}) on the set $\mathcal{N}(\mathcal{G})$. If $\mathcal{N}(\mathcal{G})=\mathcal{X}$, then the set $\mathcal{G}$ in \eqref{eq:Gamma_for_Stability} is globally asimptotically stable for the system (\ref{eq:AutoSys}). \qed
    \label{lemma:AS}\end{lemma}
    \begin{lemma} [Theorem 4.2 in \cite{bhat}]
        If there exists a continuously differentiable function $V:~\mathcal{N}(\mathcal{G})\to\R$ such that 
            \begin{align}
                &V(0)=0\text{ and }V(x)>0\tilda,\forall \tilda x\in\mathcal{N}(\mathcal{G}){\setminus}0,\label{eq:V_FTS}
                \intertext{and the following holds for some $c>0$ and $\alpha\in(0,1)$ along the trajectories of \eqref{eq:AutoSys}:}
                &\dot{V}(x)\leq -cV^\alpha,~\forall x\in\mathcal{N}(\mathcal{G})\setminus\mathcal{G},\label{eq:dotV_FTS}
            \end{align}
            then the set $\mathcal{G}$ in \eqref{eq:Gamma_for_Stability} is finite-time stable for the system (\ref{eq:AutoSys}) on the set $\mathcal{N}(\mathcal{G})$ with settling time
            $$
            t_r\leq\frac{1}{c(1-\alpha)}V^{(1-\alpha)}.
            $$
            If $\mathcal{N}(\mathcal{G})=\mathcal{X}$, then the set $\mathcal{G}$ in \eqref{eq:Gamma_for_Stability} is globally finite-time stable for the system (\ref{eq:AutoSys}).\qed
    \label{lemma:FTS}\end{lemma}
}

\section{MODELING AND PROBLEM STATEMENT}
This section provides the attitude kinematics and dynamics of rigid spacecraft. Then, it parameterise the pointing constraints  {and discusses the actuators saturation issue}. Problem statement is given in the end. 

\subsection{Attitude dynamics and kinematics}
The attitude of the spacecraft is given in terms of attitude quaternions $q=[\eta, \tilda\epsilon^T]^T\in\R^4$, specifying the orientation of the body frame $\mathcal{B}$ (attached to the spacecraft and centered in its centre of the mass) w.r.t. the inertial frame $\mathcal{I}$ (Fig. \ref{fig:scheme1}).
Then, the angular rate $\omega\in\R^3$ expressed in $\mathcal{B}$ frame describes the rotational motion of the spacecraft w.r.t. $\mathcal{I}$. Therefore, the attitude kinematics and dynamics of the spacecraft actuated by RWs is as follows:
\begin{equation}\begin{cases}
    \Dot{\eta}&=-\frac{1}{2}\omega^T{\epsilon}\\
    \Dot{\epsilon}&=\frac{1}{2}\left(\eta\omega-\omega^\times \epsilon\right)\\
    \Dot{\omega}&=I^{-1}\left(-\omega^\times H+\tau+d\right)
\end{cases},\label{eq:dynkinsys}\end{equation}
where $I\in\R^{3\times3}$ is the inertia matrix of the spacecraft  {expressed in the body frame $\mathcal{B}$ and defined relative to the spacecraft center of mass}, $d\in\R^3$ is the external disturbance torque due to space environment, $H=I\omega+h\in\R^3$ is the total angular momentum of the system, and $h,~\tau\in\R^3$ are the angular momentum stored into the wheels and the control torque developed by the electric motors respectively. They are expressed in $\mathcal{B}$ and linked by $\tau=-\Dot{h}$. Moreover, $\tau$ is related to the SMC output $u\in\R^3$ by $\tau=I^*u$, where $I^*$ is the nominal value of the inertia matrix as explained below. 

\subsubsection{Assumptions}\label{subsec:Assumption}
This  {paper} assumes that the inertia matrix of the spacecraft $I$ consists of the sum of a nominal (known) matrix $I^*\in\RR$ and uncertain (unknown) matrix $\Delta_I\in\RR$. Furthermore, it is assumed that $\Delta_I$ is bounded as $-\overline{\Delta_I}<\Delta_I<\overline{\Delta_I}$, with known $\overline{\Delta_I}\in\RR$ consisting of positive elements. Therefore, the uncertain inertia matrix $I$ satisfies $\underline{I}<I<\overline{I}$, with $\underline{I}=I^*-\overline{\Delta_I}$ and $\overline{I}=I^*+\overline{\Delta_I}$. Of course, $\overline{\Delta_I}<I^*$ to preserve the physical meanings of the inertia matrix.
Also, the Matrix inversion Lemma \cite{6189071} is used to write $I^{-1}=\left(I^*+\Delta_{I}\right)^{-1}=I^{*^{-1}}+\Hat{\Delta_{I}}$, with $\Hat{\Delta_{I}}=-I^{*^{-1}}\Delta_{I}\left(\mathbb{I}_3+I^{*^{-1}}\Delta_{I}\right)^{-1}I^{*^{-1}}$. Here, the assumption $-\Delta_I<\Delta_I<\Delta_I$ allows to define a bound $\overline{\hat{\Delta_I}}>0$ such that $-\overline{\hat{\Delta_I}}<\Hat{\Delta_I}<\overline{\hat{\Delta_I}}$.
Moreover, the non-conservative disturbance torques $d$ are caused by physical phenomena making them intrinsically continuous and bounded \cite{de2012spacecraft_dist}. Therefore, an upper bound $\overline{d}$ such that $\|d\|<\overline{d}$ is assumed to be known. Usually, $\overline{d}$ can be estimated based on the features of both the spacecraft and the orbit. Also, it is assumed that $d=d_1+d_2$, where $d_1\in\R^3$ are secular non-conservative torques and $d_2\in\R^3$ are sinusoidal disturbances with period corresponding to the orbit period. Also, the disturbance torques are assumed to satisfy $\|d_1\|<\overline{d_1}$, $\|d_2\|<\overline{d_2}$, and $\overline{d_1}<<\overline{d_2}$, while the attitude maneuver times are assumed to be shorter than one orbital period. Therefore, it can legitimately be regarded that the maximum amount of angular momentum injected into the system by the external disturbance is found after a quarter of an orbital period due to the sinusoidal behaviour of $d_2$.  
Furthermore, both the spacecraft and the RWs are assumed to have zero-angular momentum at the beginning of the control process. Reaction-type actuators capable of dampening the angular momentum of the system are always incorporated in the real spacecraft, which makes the starting condition feasible and justifies the hypothesis.

\subsection{Actuation system}
\begin{figure}
\centering
    \includegraphics[width=.6\linewidth]{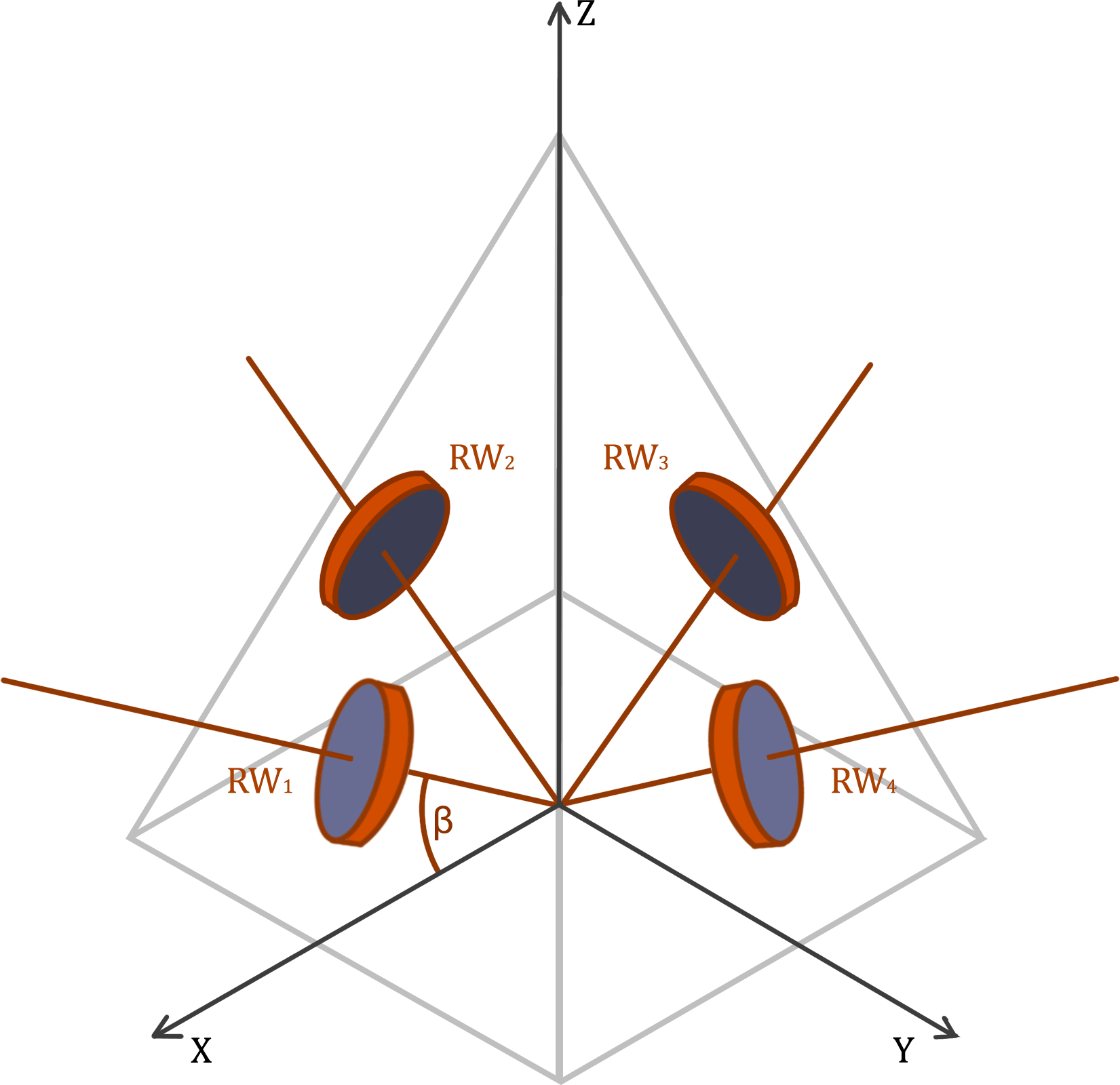}
    \caption{Pyramidal cluster of RWs \cite{app13106026}}
    \label{fig:RWs}
\end{figure}

In this paper, the actuators for attitude control consist of four RWs organized in a pyramidal configuration of the type shown in Fig. \ref{fig:RWs}. The orientation of the wheels' spin axes w.r.t. body axes are described by two angles: $\beta$ is the angle w.r.t. the base of the pyramid, while $\alpha$ is the angle between the projection of the spin axis of the first wheel in the pyramid base and the $x$ body axis ($\alpha=0$ in Fig. \ref{fig:RWs}). Therefore, $\alpha$ and $\beta$ dictate the overall torque and momentum distribution of RWs in the spacecraft body axes.
Now, consider the wheels' angular momentum vector $h^w = [h^w_{1},\dots,h^w_{4}]^T \in \mathbb{R}^4$ and the wheels' torque vector $\tau^w = [\tau^w_1,\dots,\tau^w_4]^T \in \mathbb{R}^4$, both of them expressed along the wheels' spin axes. Furthermore, take $h$ and $\tau$ as the wheels' angular momentum and torque, respectively, expressed in body frame as in Eq. (\ref{eq:dynkinsys}). Then, the following relation links $h^w$ with $h$ (and $\tau^w$ with $\tau$):
\begin{equation}
\begin{split}
    &h=Zh^w\\
    &\tau=Z\tau^w\quad 
\end{split},\quad
    Z=\begin{bmatrix}
        c_\alpha c_\beta & -s_\alpha c_\beta & -c_\alpha c_\beta & s_\alpha c_\beta\\
        s_\alpha c_\beta & c_\alpha c_\beta & -s_\alpha c_\beta & -c_\alpha c_\beta\\
        s_\beta &  s_\beta &  s_\beta &  s_\beta
    \end{bmatrix},
\label{eq:wheel2body}\end{equation}
where $s_\alpha=sin(\alpha)$ and $c_\alpha=cos(\alpha)$. Since $Z\in\mathbb{R}^{3\times4}$, in the following the inverse of $Z$ is understood in the sense of the Moore-Penrose pseudo-inverse matrix $Z^+$.

It is well known that RWs suffer from the following saturation constraints: $|\tau^w_i|\leq\overline{\tau^w}$, $|h^w_{i}|\leq\overline{h^w}$, with $\overline{\tau^w}$, $\overline{h^w}$ representing the maximum torque and the maximum angular momentum developed by each wheel, respectively. 
Consequently, the maximum angular momentum and the maximum torque developed by the pyramidal cluster of RWs are bounded by $\overline{H}$ and $\overline{\tau}$ as follows:
\begin{equation}
    \|h\|\leq\overline{H},\quad \|\tau\|\leq\overline{\tau}.    \label{eq:act_const}
\end{equation}
$\overline{H}$ and $\overline{\tau}$ can be obtained with the same procedure, which is well detailed in \cite{YOON2014109} and requires the saturation values of each wheel and the geometry of the cluster to be known. As an example, below is the procedure for obtaining $\overline{H}$.
\begin{figure}
\centering
    \includegraphics[width=.6\columnwidth]{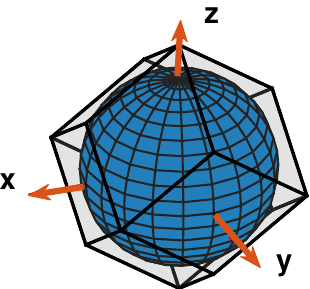}
    \caption{Momentum (torque) sphere inscribed in the polyhedron of the momentum (torque) envelope.}
    \label{fig:envelope}
\end{figure}
The first step is to extract the polyhedron in Fig. \ref{fig:envelope} corresponding to the momentum envelope, i.e. the capacity of the cluster of RWs to store angular momentum.
A single facet of the polyhedron is obtained by following the two steps below.\begin{enumerate}[(I)]
    \item The angular momentum of two wheels is fixed at the saturation values $\pm\overline{h^w}$ and the other two wheels are left free to vary, assuming all possible values within the domain bounded by the saturation values.
    \item The angular momentum in wheels' frame is transformed in $\mathcal{B}$ through Eq. (\ref{eq:wheel2body}).
\end{enumerate} 
The other facets of the polyhedron are then obtained by repeating steps (I) and (II) with all combinations of two saturated wheels and two free wheels. Finally, $\overline{H}$ is the radius of the sphere inscribed in the polyhedron, i.e. the momentum sphere. In other words, the momentum sphere is defined as the locus of the angular momentum vector of the cluster with magnitude lower than $\overline{H}$. Since the momentum sphere of radius $\overline{H}$ is completely enclosed in the momentum envelope, satisfying the first of the inequalities (\ref{eq:act_const}) ensures that $\left|h_i^w\right|\leq\overline{h}$ $\forall~i=1,\dots,4$. By repeating the procedure with the torque values $\tau$ and the corresponding saturation values $\pm\overline{\tau^w}$, the polyhedron corresponding to the torque envelope and the inscribed sphere of radius $\overline{\tau}$ corresponding to the torque sphere are obtained. Hence, satisfying the second of the inequalities (\ref{eq:act_const}) avoids the torque saturation of the wheels.

In particular, the constraint $\|h\|\leq\overline{H}$ implies that the attitude maneuver speed has to be kept bounded by some positive constant $\overline{\omega}$, i.e. $\|\omega\|\leq\overline{\omega}$ to preserve the controllability of the satellite.   
Indeed, if a reaction wheel is saturated, it can not accelerate further, which means that the spacecraft can not be controlled around the axis of the saturated wheel. In order to determine $\overline{\omega}$, consider the accumulation of angular momentum over a quarter of the orbital period due to the disturbance torque: $h_d=\int_0^{\nicefrac{T}{4}} d(t) dt$, where $T$ is the orbital period and $\nicefrac{T}{4}$ is considered by virtue of the sinusoidal behaviour of $d_2$. 
Consequently,  $h=h_d-I\omega$ holds true at any time instant, and exploiting the properties of the $l^2-$norm together with $I<\overline{I}$ it follows: 
\begin{equation}\begin{split}
    \|h\|\leq\|h_d\|+\|I\omega\|~&\Rightarrow~\|\omega\|\leq\overline{\omega}=\frac{\overline{H}-\overline{h_d}}{\|\overline{I}\|}
\end{split}\label{eq:MAXom}\end{equation}
where the last inequality in equation above identifies the necessary condition on $\omega$ to avoid angular momentum saturation while maneuvering, while $\overline{h_d}=(\overline{d_1}+0.707\overline{d_2})\frac{T}{4}$ due to the sinusoidal behaviour of $d_2$.

\subsection{Pointing constraints}\label{subsec:PointingConst}
\begin{figure}[tbp]
\centering
\includegraphics[width=0.7\columnwidth]{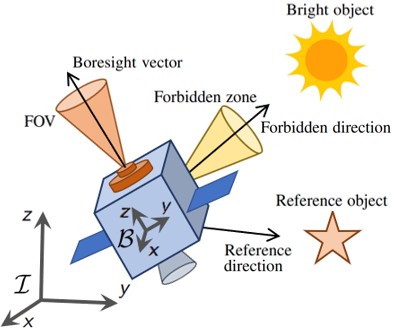}
    \vspace{0ex}
    \caption{Spacecraft scheme with pointing constraint.}
    \label{fig:scheme1}
\end{figure}

Pointing constraints are introduced to prevent bright objects from entering the sensor Field Of View (FOV) during spacecraft reorientation, whose objective is to align the  boresight vector of the on-board instrument along the reference direction pointing toward the scientific object (Fig. \ref{fig:scheme1}). Referring again to Fig. \ref{fig:scheme1}, $\hat{m}$ is the unit vector aligned with the boresight vector, $\Hat{n}_j$ is the unit vector aligned with the forbidden direction pointing towards the $j-$th celestial body, and $\theta_j=ang(\hat{m},\hat{n}_j)$. 
Then, for each bright object a forbidden zone is defined as a cone with the axis along $\Hat{n}_j$ (forbidden direction) and a half apex angle $\underline{\theta}>0$. The latter is defined such that if $\left|\theta_j\right|\geq\underline{\theta}$, then the unwanted object stays outside the sensor FOV. Now, $\theta_j$ is put into relation with $\Hat{m},\tilda\Hat{n}_j$ to formulate the pointing constraints. First, it is reasonable to assume $\Hat{m}$ to be known in $\mathcal{B}$ coordinates and $\Hat{n}_j$ to be known in $\mathcal{I}$ coordinates. Also, $\Hat{m}_I$ in $\mathcal{I}$ coordinates can be obtained through the rotation matrix {$R(q)=(\eta^2+\epsilon^T\epsilon)\mathbb{I}_3+2\epsilon\epsilon^T-2\eta\epsilon^\times$} built with the quaternions describing the orientation of $\mathcal{B}$ w.r.t. $\mathcal{I}$: $\Hat{m}_I=R^T\Hat{m}$. This allows to derive the attitude quaternions $\Tilde{q}_j=[\Tilde{\eta}_j\tilda\Tilde{\epsilon}_j^T]^T$ describing the shortest rotation between $\Hat{m}$ and $\Hat{n}_j$:
\begin{equation}\begin{split}
    a_j&=\Hat{m}_I^T\Hat{n}_j\\
    b_j&{=\Hat{m}_I^\times\Hat{n}_j} 
\end{split},\tilda
    \Tilde{\eta}_j=\frac{1 + a_j}{\|[1 + a_j, \tilda b_j^T]\|},\tilda \Tilde{\epsilon}_j^T=\frac{b_j^T}{\|[1 + a_j, \tilda b_j^T]\|}.
    \label{eq:tilde_q}
\end{equation}
$\tilde{q}_j$ will later be employed in the formulation of the repulsive layer of the APF. At this stage, however, only $\tilde{\eta}_j$ is utilized to express the pointing constraint. Since $\tilde{q}_j$ represents the minimal angular path between the vectors $\hat{m}$ and $\hat{n}_j$, it follows that $\tilde{\eta}_j = \cos(\theta_j/2)$.
Therefore, $\left|\Tilde{\eta}_j\right|\leq \Tilde{\underline{\eta}}=cos(\underline{\theta}/2)$ implies $\left|\theta_j\right|\geq\underline{\theta}$ (bright object outside the sensor FOV). The properties of attitude quaternions yield $\|\Tilde{\epsilon}_j\|=\sqrt{1-\Tilde{\eta_j}^2}$, allowing to define $\underline{\Tilde{\epsilon}}=\sqrt{1-\Tilde{\underline{\eta}}^2}$. Therefore, the pointing constraint $\left|\theta_j\right|\geq\underline{\theta}$ can be expressed with the vectorial part of $\Tilde{q}_j$ as follows: 
\begin{equation}
    \|\Tilde{\epsilon}_j\|\geq\underline{\Tilde{\epsilon}},\quad\underline{\Tilde{\epsilon}}=\sqrt{1-\text{cos}^2\frac{\underline{\theta}}{2}}
    \label{eq:underline_eps}
\end{equation}

\subsection{Problem statement}
The objective is to design a guidance and control strategy such that $q\to q_d$ and $\omega\to \mathbf{0}_3$, where $q_d$ denotes the desired attitude that guarantees the alignment of the boresight vector to the reference direction.  
In addition, it is required that the guidance and control strategy keeps bright objects outside the sensor FOV while performing the attitude maneuver, i.e. that it guarantees $\|\Tilde{\epsilon}_j\|\geq\Tilde{\underline{\epsilon}}\tilda\forall\tilda j$ as discussed above.
Finally, the attitude control law has to comply with the actuators saturations, thus providing control signal which guarantees $\|h\|\leq\overline{h}$, $\|\tau\|\leq\overline{\tau}$ while performing the attitude maneuver, at least for initial conditions $\omega_0=h_0=\mathbf{0}_3$.

\section{GUIDANCE AND CONTROL STRATEGIES}
In this article, a combined guidance and control approach is adopted to address the attitude control problem. The guidance algorithm is based on APF and generates the reference angular rate $\omega^*$ which describes an attitude maneuver guaranteeing that $q\to q_d$ and $\omega\to\mathbf{0}_3$ are achieved together with $\|\Tilde{\epsilon}_j\|\geq\underline{\Tilde{\epsilon}}$. Therefore, if the attitude of the spacecraft evolves with $\omega=\omega^*$ the spacecraft achieves the desired orientation $q_d$ while ensuring that the boresight vector successfully avoids the forbidden zones for the whole duration of the attitude maneuver. Then, the control algorithm is based on SMC theory and is responsible for achieving $\omega\to\omega^*$. Therefore, the SMC computes the control input $u$ such that the control torque  {$\tau=I^*u$} steers the spacecraft along the reference angular path generated by the APF (Fig. \ref{fig:orbital_simu}).

\subsection{Guidance strategy}
Here, the fully quaternion-based APF guidance strategy is established to find the angular path that steers the spacecraft attitude toward the reference avoiding forbidden zones. The angular speed field $\omega^*$ is evaluated using (i) the quaternions describing the shortest rotation from actual to desired orientation and (ii) the quaternions describing the shortest rotation leading the boresight vector to the forbidden direction. 
Furthermore, quaternions at points (i) and (ii) are built by using data concerning the state vector of the spacecraft $(q,\tilda\omega)$, its geometry $(\Hat{m})$, the environment in which it moves $(\Hat{n}_j)$ and the desired final orientation $(q_d)$. Then, the reference quaternions $q^*$ is obtained by propagating the actual quaternions $q$ with $\omega^*$. This strategy assigns to the desired final orientation an attractive potential field and a repulsive one related to each forbidden zone.

\subsubsection{Attractive APF}
The spacecraft approaches the goal thanks to an attractive field $U_a$ built by combining parabolic and conic functions as follows:
\begin{equation}\begin{split}
    U_{a}=
    \begin{cases}
   \frac{\alpha_1}{2}\|\epsilon_e\|^{2}\tilda\text{sign}_+(\eta_e)  &\text{if }\tilda \|\epsilon_e\|\leq \overline{\epsilon_e} \\
   \frac{\alpha_2}{2}\|\epsilon_e\|\tilda\text{sign}_+(\eta_e)  &\text{if }\tilda \|\epsilon_e\|> \overline{\epsilon_e} 
   \end{cases} ,
   \label{eq:Ua}
\end{split}\end{equation}
where $\eta_e,\tilda\epsilon_e$ form the quaternion error $q_e=[\eta_e\tilda\epsilon_e^T]=\overline{q_d}\otimes q$. 
For  $\|\epsilon_e\|=\overline{\epsilon_e}$  the potential changes shape from conical $(\|\epsilon_e\|> \overline{\epsilon_e})$ to parabolic $(\|\epsilon_e\|\leq \overline{\epsilon_e})$.
The relative attractive angular rate $\omega_a$ is obtained through the gradient of $U_a$, as follows:
\begin{equation}\begin{split}
    \omega_{a}=\nabla_{\epsilon_e}U_a=
    \begin{cases}
   {\alpha_1}\epsilon_e\tilda\text{sign}_+(\eta_e)  &\text{if }\tilda \|\epsilon_e\|\leq \overline{\epsilon_e} \\
   {\alpha_2}\frac{\epsilon_e}{\|\epsilon_e\|}\tilda\text{sign}_+(\eta_e)  &\text{if }\tilda \|\epsilon_e\|> \overline{\epsilon_e}
   \end{cases} ,
   \label{eq:omega_a}
\end{split}\end{equation} 
The chosen field $U_a$ combines the advantages of parabolic and conic functions:  (i)  $\omega_a\to0$ for $\epsilon_e\to0$, i.e. for $q\to q_d$ and (ii)  $\|\omega_a\|=\alpha_2=const.$ for high pointing errors. The property (i) is given by the parabolic function and allows to avoid any singularities for $\epsilon_e=0$. Instead, the property (ii) is given by the conic function and allows to saturate the angular rate of the spacecraft, thus avoiding actuators saturations for appropriately chosen $\alpha_2$. Keeping in mind that the $h-$saturation of RWs is avoided if $\|\omega\|\leq\overline{\omega}$, it is selected $\alpha_2=\overline{\omega}/2$. 
Moreover, $\alpha_1=\alpha_2/\overline{\epsilon_e}$ is selected to avoid $\omega_a-$discontinuities at $\|\epsilon_e\|=\overline{\epsilon_e}$. 
Finally, the threshold value
\begin{equation}
    \overline{\epsilon_e}=\frac{(\overline{\omega}/2)^2}{\|\overline{I}^{-1}\|\overline{\tau}}
    \label{eq:overline_eps}
\end{equation}
accounts for actuators limitations, being defined as the angular distance the spacecraft covers to stop its angular motion with initial velocity $\|\omega\|=\overline{\omega}/2$ and maximum constant deceleration $\|\overline{I}^{-1}\|\overline{\tau}$ (the equations of circular motion with constant tangential acceleration are considered to compute $\overline{\epsilon_e}$). 

\subsubsection{Repulsive APF} 
An hyperbolic potential $U_{r,j}$ is built to shape each forbidden zone, which results in a repulsive angular speed $\omega_r=\sum_{j=1}^N\omega_{r,j}$ ($N$ number of bright objects) leading the satellite to rotate the boresight vector away from each forbidden direction. 
The true and the minimum safe angular distance between the boresight vector and the forbidden direction are specified by $\|\Tilde{\epsilon}_j\|$ and $\underline{\Tilde{\epsilon}}$ respectively, as derived in Section \ref{subsec:PointingConst}. Consequently, $\omega_{r,j}$ is obtained from the gradient of $U_{r,j}$ as below. The selection of $\zeta$ is discussed in Section \ref{subsub:REFTRAJ}.
\begin{equation}
    U_{r,j}=\zeta\frac{1}{\|\Tilde{\epsilon}_j\|},\quad\omega_{r,j}=\nabla_{\Tilde{\epsilon}_j}U_{r,j}= {-}\frac{\zeta}{\|\Tilde{\epsilon}_j\|^3}\Tilde{\epsilon}_j,\quad \zeta=\alpha_2\underline{\Tilde{\epsilon}}^2
    \label{eq:Ur}
\end{equation}

\subsubsection{Reference trajectory}\label{subsub:REFTRAJ}
The reference angular rate $\omega^*$ is obtained by combining both attractive (Eq. \ref{eq:omega_a}) and repulsive (Eq. \ref{eq:Ur}) layers as follows:
\begin{equation}
    \omega^*=R(q)\left(\omega_a+\sum_{j=1}^N\omega_{r,j}\right),
    \label{eq:omega^*}
\end{equation}
where $R(q)$  allows to obtain $\omega^*$ in $\mathcal{B}$ coordinates. 
Now, the following Proposition describes the reference trajectory of the spacecraft rotational motion resulting from $\omega^*$.
\begin{proposition}
    Take the reference angular rate $\omega^*$ in Eq. (\ref{eq:omega^*}), with $\omega_a$ and $\omega_{r,j}$ as in Eqs. (\ref{eq:omega_a}) and (\ref{eq:Ur}), respectively, and $\alpha_2=\overline{\omega}/2$, $\alpha_1=\alpha_2/\overline{\epsilon_e}$, $\zeta=\alpha_2\underline{\Tilde{\epsilon}}^2$, $\overline{\omega}$ as in Eq. (\ref{eq:MAXom}), $\overline{\epsilon_e}$ as in Eq. (\ref{eq:overline_eps}), and $\underline{\Tilde{\epsilon}}$ as in Eq. (\ref{eq:underline_eps}).  Assume now that the forbidden directions are sufficiently spaced apart so that the repulsive action is imparted by a single forbidden body at every point on the celestial sphere. Assume also that $\|\Tilde{\epsilon}_j\|>\underline{\epsilon}$  is satisfied  $\forall\tilda j=1,\dots,N$ by the initial orientation of the spacecraft $q(t=0)=q_0=[\eta_0,\epsilon_0^T]$.  
    Finally, let us recall the assumptions of Section \ref{subsec:Assumption} on the null angular momentum at $t=0$ and the boundedness of disturbances and uncertainties.
    
    Then, $\omega^*$ describes an attitude maneuver which achieves $q\to q_d$, i.e. $\epsilon_e\to\mathbf{0}_3$, while fulfilling $\|\Tilde{\epsilon}_j\|>\underline{\Tilde{\epsilon}}$ at any time of the control process. Also, $\omega^*$ avoids angular momentum saturation guaranteeing $\|h\|<\overline{H}$ during the attitude maneuver.

    \begin{proof} \textit{of Proposition \ref{prop}}
    
    Let us first analyse a scenario in which there are no forbidden zones affecting the attitude maneuver. In that case, Eq. (\ref{eq:omega^*}) yields $\omega^*=R(q)\omega_a$, with $\omega_a$ as in Eq. (\ref{eq:omega_a}) leading  leading the spacecraft to the orientation $q_d$ via the shortest rotation from any initial condition $q_0$.    
    Indeed, Eq. (\ref{eq:omega_a}) reveals that $\omega_a$ is directed along the eigenaxis describing the shortest rotation between the true and desired attitude. Furthermore, the function ${sgn}_+(\eta_e)$ successfully avoids the unwinding phenomenon related to quaternions’ dual coverage, ensuring that the shortest angular path is taken to complete the maneuver. 
    Also, Eqs. (\ref{eq:omega_a}), (\ref{eq:omega^*}) yield $\|\omega^*\|\leq\overline{\omega}/2$, which fulfills Eq. (\ref{eq:MAXom}) thus avoiding angular momentum saturation and preserving the controllability of the spacecraft under the validity assumptions of the Proposition. Therefore, APF-based guidance returns an angular velocity profile $\omega^*$ that can be reached by the satellite and that cancels the attitude error if no forbidden zones impact the guidance output.

    Let us now consider the satellite oriented close to a single forbidden zone $j$ so that $\omega_r=\omega_{r,j}$ and
    \begin{equation}
        \|\omega_{r,j}\|=\left(\frac{\underline{\Tilde{\epsilon}}}{\|\Tilde{\epsilon}_j\|}\right)^2\overline{\omega}/2
        \label{eq:norm_om_r}
    \end{equation} 
    due to Eq. (\ref{eq:Ur}) together with the above selection of $\zeta$ and $\alpha_2$. In that case, Eq. (\ref{eq:omega^*}) reveals that $\omega^*$ is given by the sum of two contributions, whose comparative strength depends on the value $\|\Tilde{\epsilon_j}\|$ as follows:
    \begin{subequations}\begin{align}
        \|\omega_{r}\|=\|\omega_{r,j}\|>\|\omega_a\| \tilda &\text{if} \tilda \|\Tilde{\epsilon}_j\|<\underline{\Tilde{\epsilon}}\label{eq:a}\\
        \|\omega_{r}\|=\|\omega_{r,j}\|=\|\omega_a\| \tilda &\text{if} \tilda \|\Tilde{\epsilon}_j\|=\underline{\Tilde{\epsilon}}\label{eq:b}\\
        \|\omega_{r}\|=\|\omega_{r,j}\|<\|\omega_a\| \tilda &\text{if} \tilda \|\Tilde{\epsilon}_j\|>\underline{\Tilde{\epsilon}}\label{eq:c}
\end{align}\label{eq:om_r_vs_om_a}\end{subequations}
In order to justify Eq. (\ref{eq:om_r_vs_om_a}), let us recall that $\|\omega_a\|\leq\overline{\omega}/2$ due to Eq. (\ref{eq:omega_a}) and the proposed selection of $\alpha_1$ and $\alpha_2$. Furthermore, $\|\omega_r\|\lesseqgtr\overline{\omega}/2$ depending on $\nicefrac{\underline{\Tilde{\epsilon}}}{\|\Tilde{\epsilon}_j\|}\lesseqgtr1$ according to Eq. (\ref{eq:norm_om_r}). 
Consequently, Eq. (\ref{eq:om_r_vs_om_a}) reveals that the behaviour of the APF-based guidance  is as follows. 
If the boresight vector is at the borders of a forbidden zone, i.e. $\tilda \|\Tilde{\epsilon}_j\|=\underline{\Tilde{\epsilon}}$, the APF guidance yields $\|\omega_r\|=\|\omega_a\|$ to avoid further approach. Furthermore, if the boresight vector is already inside the forbidden zone, i.e. $\tilda \|\Tilde{\epsilon}_j\|<\underline{\Tilde{\epsilon}}$, the repulsive layer of APF takes absolute priority over the attractive part and the forbidden zone is disengaged. However, due to Eq. (\ref{eq:b}) this situation never arises if $q_0$ fulfills the validity assumptions of the Proposition. Then, $\|\omega_r\|<\|\omega_a\|$ in the proximity of the forbidden zone. Here, $\omega^*$ is globally directed towards $q_d$ while the crossing of the forbidden zones is avoided thanks to  $\|\omega_r\|\to\|\omega_a\|$ as $\|\Tilde{\epsilon}_j\|\to\underline{\Tilde{\epsilon}}$. 
Also, Eqs. (\ref{eq:b}), (\ref{eq:c}) yield $\|\omega^*\|\leq\|\omega_a\|+\|\omega_{r}\|\leq\overline{\omega}$, while the condition in Eq. (\ref{eq:a}) never arises as explained above. Therefore, if the satellite is oriented close to a forbidden zone, the APF-based guidance returns a reachable $\omega^*$ such that the satellite safely moves beyond the influence zone of the forbidden zone until recovers the behaviour described in the first case analysed, i.e. $\omega^*R(q)=\omega_a$. 
    \end{proof}
\label{prop}\end{proposition}
    
\begin{figure}
        \centering
        \includegraphics[width=0.9\linewidth]{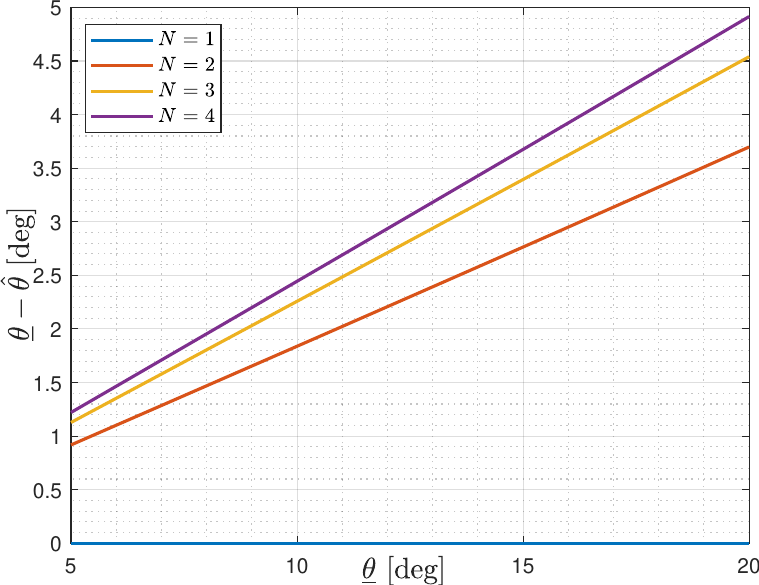}
        \caption{ {Decrease of $\hat\theta$ with respect to $\underline{\theta}$ due to different number of forbidden zones $N$.}}
        \label{fig:theta_hat}
    \end{figure}

\begin{figure}
        \centering
        \includegraphics[width=0.9\linewidth]{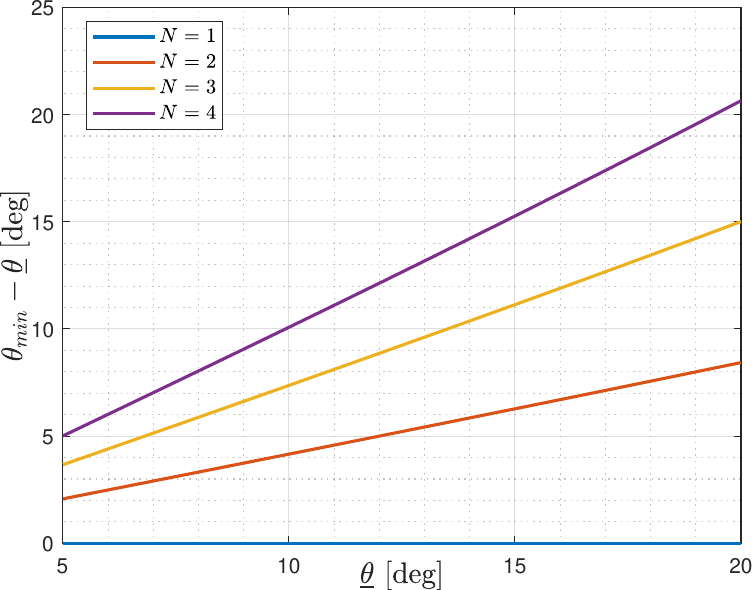}
        \caption{ {Increase of $\theta_{\text{min}}$ with respect to $\underline{\theta}$ due to different number of forbidden zones $N$.}}
        \label{fig:theta_min}
    \end{figure}

\begin{figure}
        \centering
        \includegraphics[width=0.9\linewidth]{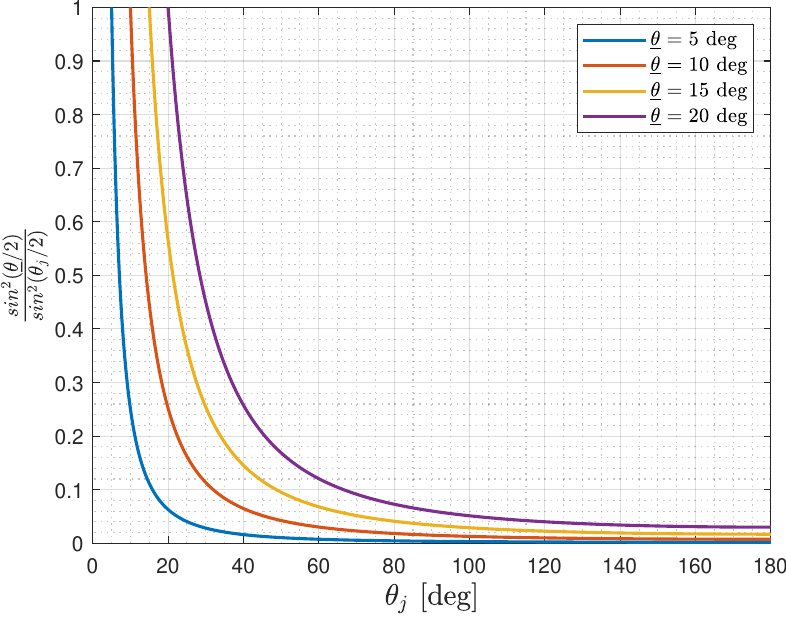}
        \caption{ {Decay of the contribution of a forbidden zone $j$ to the total potential with increasing $\theta_j$ and for different values of $\underline{\theta}$.}}
        \label{fig:decay}
    \end{figure}

Proposition \ref{prop} proved to be true if the repulsive potential is generated by a single forbidden body, i.e. $N=1$. 
To conclude this Section, Corollary \ref{corol:toProp1} gives the conditions to extend some results of Proposition \ref{prop} to the case where the repulsive potential is generated by multiple forbidden zones, i.e. $N>1$.
\begin{corollary}
         {The statement of Proposition 1 is modified as follows if the spacecraft operates under the influence of multiple forbidden zones:
        \begin{enumerate}[(i)]
             \item $\omega^*$ describes an attitude maneuver which achieves $q\to q_d$ while fulfilling 
             \begin{equation}
                \left|\theta_{\underline{j}}\right|\geq\Hat{\theta}=2\text{arcsin}\left(\sqrt{\frac{\text{sin}^2\nicefrac{\underline{\theta}}{2}}{1+\sum_{\substack{j=1 \\ j\neq\underline{j}}}^N\frac{\text{sin}^2\nicefrac{\underline{\theta}}{2}}{\text{sin}^2\nicefrac{\theta_j}{2}}}}\right)~\forall\underline{j}=1,\dots,N
                 \label{eq:item1coro}
             \end{equation}
             where $\theta_{\underline{j}}$ is the angle between the forbidden zone $\underline{j}$ and the axis of the instrument, $\theta_j$ is the true angular distance between the boresight vector and any forbidden direction $j$, and $\underline{\theta}$ is the minimum safe angular distance.            \label{coro:item1}
            \item $\omega^*$ guarantees $\|h\|<\overline{H}$, provided that the following condition is satisfied:
        \begin{equation}
            \left|\theta_j\right|\geq\theta_{\text{min}}=2\text{arcsin}\left(\sqrt{N}\text{sin}\left(\frac{\underline{\theta}}{2}\right)\right)\quad\forall\tilda j=1,\dots,N
            \label{eq:corollary}
        \end{equation}
    \label{coro:item2} 
        \end{enumerate}}
    \begin{proof}\textit{of Corollary \ref{corol:toProp1}}
    
     For $N>1$ the global repulsive action is obtained by summing up the contributions of all forbidden bodies, so that the repulsive angular speed generated by a particular forbidden zone $\underline{j}$ must be put in relation with the following:
    \begin{equation}
        \left\|\omega_a+\sum_{\substack{j=1, j\neq\underline{j}}}^N\omega_{r,j}\right\|.
        \label{eq:und_j_vs_all}
    \end{equation}
    Indeed, any $\omega_{r,j}$ with $j\neq\underline{j}$ may be directed towards the forbidden zone $\underline{j}$, thus pushing the sensor axis towards the forbidden zone $\underline{j}$. The following demonstrates that $\|\omega_{r,\underline{j}}\|$ takes values greater than the output of Eq. \eqref{eq:und_j_vs_all} for $\left|\theta_{\underline{j}}\right|<\hat{\theta}$, thereby proving point (\ref{coro:item1}) of Corollary \ref{corol:toProp1}. First, Eq. \eqref{eq:und_j_vs_all} is bounded as follows due to Eqs. \eqref{eq:omega_a} and \eqref{eq:norm_om_r}:
    \begin{equation}
        \|\omega_a\|+\|\sum_{\substack{j=1 \\ j\neq\underline{j}}}^N\omega_{r,j}\|\leq\frac{\overline{\omega}}{2}+\frac{\overline{\omega}}{2}\sum_{\substack{j=1 \\ j\neq\underline{j}}}^N\left(\frac{\underline{\Tilde{\epsilon}}}{\|\Tilde{\epsilon}_j\|}\right)^2
        \label{eq:bound_undj}
    \end{equation}
    Therefore, $\omega_{r,\underline{j}}$ arrests the advancement of the instrument axis toward the bright body if the following is satisfied:
    \begin{equation}
        \|\omega_{r,\underline{j}}\|=\frac{\overline{\omega}}{2}\left(\frac{\underline{\Tilde{\epsilon}}}{\|\Tilde{\epsilon}_{\underline{j}}\|}\right)^2\geq\frac{\overline{\omega}}{2}+\frac{\overline{\omega}}{2}\sum_{\substack{j=1 \\ j\neq\underline{j}}}^N\left(\frac{\underline{\Tilde{\epsilon}}}{\|\Tilde{\epsilon}_j\|}\right)^2
        \label{eq:und_j_draw}
    \end{equation}
    Due to the properties of attitude quaternions (see Eq. \ref{eq:underline_eps}):
    \begin{equation}
        \underline{\Tilde{\epsilon}}^2=1-\text{cos}^2\frac{\underline{\theta}}{2}=\text{sin}^2\frac{\underline{\theta}}{2},\quad        \|\Tilde{\epsilon}_j\|^2=1-\text{cos}^2\frac{\theta_j}{2}=\text{sin}^2\frac{\theta_j}{2}.   
    \label{eq:eps2theta}\end{equation}
    Therefore, Eq. \eqref{eq:und_j_draw} can be written as follows:
    \begin{equation}
        \frac{\text{sin}^2\nicefrac{\underline{\theta}}{2}}{\text{sin}^2\nicefrac{\theta_{\underline{j}}}{2}}\geq1+\sum_{\substack{j=1 \\ j\neq\underline{j}}}^N\left(\frac{\text{sin}^2\nicefrac{\underline{\theta}}{2}}{\text{sin}^2\nicefrac{\theta_j}{2}}\right)^2\tilda\Leftrightarrow\tilda\text{sin}^2\frac{\theta_{\underline{j}}}{2}\leq\frac{\text{sin}^2\nicefrac{\underline{\theta}}{2}}{1+\sum_{\substack{j=1 \\ j\neq\underline{j}}}^N\left(\frac{\text{sin}^2\nicefrac{\underline{\theta}}{2}}{\text{sin}^2\nicefrac{\theta_j}{2}}\right)^2},
    \end{equation}
    which is equivalent to Eq. \eqref{eq:item1coro} thus certifying that the APF strategy realizes $\theta_{\underline{j}}\geq\hat{\theta}$ for each $\underline{j}=1,\dots,N$ at any time of the control process. As in the case $N=1$, once beyond the influence of the repulsive fields the satellite continues to rotate under the attracting field, thus realizing $q\to q_d$.

     Then, point (\ref{coro:item2}) is proved as follows. According to Eq. (4), the output of the APF algorithm realizes $\|h\|<\overline{H}$ if it satisfies $\|\omega^*\|\leq\overline{\omega}$. Also, inserting Eqs. (8) and (12) into Eq. (11) yields
    \begin{equation}
        \|\omega^*\|\leq\|\omega_a\|+\|\sum_{j=1}^N\omega_{r,j}\|\leq\frac{\overline{\omega}}{2}+\sum_{j=1}^N\left(\frac{\underline{\Tilde{\epsilon}}}{\|\Tilde{\epsilon}_j\|}\right)^2\frac{\overline{\omega}}{2}.
        \label{eq:norm_om^*_tot}
    \end{equation}
    According to Eqs. \eqref{eq:norm_om^*_tot} and \eqref{eq:eps2theta}, $\|\omega^*\|\leq\overline{\omega}$ is ensured if the following is satisfied:
    \begin{equation}
        \frac{\overline{\omega}}{2}+\frac{\overline{\omega}}{2}\text{sin}^2\frac{\underline{\theta}}{2}\sum_{j=1}^N\frac{1}{\text{sin}^2\frac{\theta_j}{2}}\leq\overline{\omega}~\Leftrightarrow~\text{sin}^2\frac{\underline{\theta}}{2}\sum_{j=1}^N\frac{1}{\text{sin}^2\frac{\theta_j}{2}}\leq1.
        \label{eq:eq3}
    \end{equation}
    Also, condition \eqref{eq:corollary} is equivalent to
    \begin{equation}
        \sum_{j=1}^N\frac{1}{\text{sin}^2\frac{\theta_j}{2}}\leq\frac{1}{N\text{sin}^2\frac{{\underline{\theta}}}{2}},
    \end{equation}
    which satisfies the inequality \eqref{eq:eq3} as showed below:
    \begin{equation}
        \text{sin}^2\frac{\underline{\theta}}{2}\sum_{j=1}^N\frac{1}{\text{sin}^2\frac{\theta_j}{2}}\leq\text{sin}^2\frac{\underline{\theta}}{2}\frac{1}{N\text{sin}^2\frac{{\underline{\theta}}}{2}}\leq 1
    \end{equation}
    due to $N\geq1$, and the proof of Corollary \ref{corol:toProp1} is complete.
    \end{proof}
    \label{corol:toProp1}\end{corollary}
    Corollary \ref{corol:toProp1} discusses the robustness of Proposition 1     respect to the assumption that at any point on the celestial sphere the repulsive action is exerted by a single forbidden zone. Also, it gives some conditions which allows to extend the validity of Proposition 1 to the case when the repulsive field is generated by the superposition of multiple forbidden zones.
    Furthermore, Eqs. \eqref{eq:item1coro} and (\ref{eq:corollary}) reveal that $\hat{\theta}=\theta_{\text{min}}=\underline{\theta}$ for $N=1$, while $N>1$ implies $\hat{\theta}<\underline{\theta}$ and $\theta_{\text{min}}>\underline{\theta}$. 
    Both $(\underline{\theta}-\hat{\theta})$ and $(\theta_{\text{min}}-\underline{\theta})$ increases with $N$ as shown in Figures \ref{fig:theta_hat} and  \ref{fig:theta_min}, respectively.  They consider $\hat{\theta}$ in Eq. \eqref{eq:item1coro}, $\theta_{\text{min}}$ in Eq. \eqref{eq:corollary}, $N=\{1,2,3,4\}$ and $\underline{\theta}\in[5,~20]$ deg. Furthermore, $\hat{\theta}$ in Fig. \ref{fig:theta_hat} is computed assuming $\theta_j=\theta_{\text{min}}$ for any  $j=1,\dots,N$ such that $j\neq\underline{j}$.
    \begin{remark}
    Note that point (\ref{coro:item1}) of Corollary \ref{corol:toProp1} is highly conservative, as it is derived from Eq. \eqref{eq:und_j_draw}. In practice, it is sufficient for $\|\omega_{r,\underline{j}}\|$ to exceed the output of Eq. \eqref{eq:und_j_vs_all} for the satellite to disengage from the forbidden zone $\underline{j}$.
        Also, the validity condition of point (\ref{coro:item2}) of Corollary \ref{corol:toProp1} (Eq. \ref{eq:corollary}) is sufficient but not necessary for ensuring $\|h\|<\overline{H}$. Even if $|\theta_j|<\theta_\text{min}$ for some $j=1,\dots,\overline{j}$ ($\overline{j}<N$), the APF algorithm can still achieve $\|\omega^*\|\leq\overline{\omega}$, i.e. $\|h\|<\overline{H}$, provided that the remaining $N-\overline{j}$ values of $\theta_j$ satisfy Eq. \eqref{eq:eq3}.
    \label{remark1}\end{remark}
    Remark \ref{remark1} holds practical relevance due to the exponential decay of the repulsive field, whereby any single term of the summation in Eqs. \eqref{eq:item1coro} or \eqref{eq:eq3} decrease steeply with $\theta_j$. This behaviour is clearly evidenced by Fig. \ref{fig:decay}, which shows the trend of $sin^2\left(\nicefrac{\underline{\theta}}{2}\right)/sin^2\left(\nicefrac{\theta_j}{2}\right)$ against $\theta_j$ for different values of $\underline{\theta}$.

Finally, the reference quaternions $q^*=[\eta^*,\tilda\epsilon^*]$ are 
computed by integrating the equations below with $\omega^*$ from Eq. (\ref{eq:omega^*}) and initial conditions $q^*_0=q_0$.
\begin{equation}
   \Dot{\eta}^*=-\frac{1}{2}\omega^{*^T}{\epsilon^*}, \quad \Dot{\epsilon}^*=\frac{1}{2}\left(\eta^*\omega^*-\omega^{*\times} \epsilon^*\right)
\label{eq:q^*}\end{equation}

\subsection{Control strategy}
This section details the design of the SMC algorithm, which is responsible for tracking the reference attitude trajectory computed by the APF. 
Indeed, manipulating the attitude trajectories in such a way that $(q,\omega)\to(q^*,\omega^*)$ guarantees the resolution of the problem statement as stated in Proposition 1. 
As briefly mentioned in the Introduction, the design process of a SMC consists of defining a control law to confine the closed-loop dynamics on the sliding surface.

The sliding variable $\sigma\in\R^3$ is defined based on the linear combination of the states errors $e_\omega\in\R^3$, {$e_\epsilon\in\R^3$} as follows:
\begin{align}
    e_\omega=&\omega-\omega^*,\tilda e_q=\overline{q^*}\otimes q,\label{eq:errors}\\
    \sigma=&e_\omega+\Lambda e_\epsilon,\tilda \Lambda=\text{diag}\left(\lambda\mathbf{1}_3\right),~\lambda\in\R^+,\label{eq:sigma}\\
\intertext{where $e_q=[e_\eta\tilda e_\epsilon^T]^T$ ($e_\eta\in\R$ {is the scalar part} and $e_\epsilon\in\R^3$ {is the vectorial part}) is the attitude quaternion error and $\Lambda\in\R^{3\times3}$ is diagonal positive definite matrix with the vector $\lambda\mathbf{1}_3$ on the diagonal.   
The sliding surface is the region where the sliding variable $\sigma$ (Eq. \ref{eq:sigma}) is null. {It is important to note that the control input explicitly appears in the first time derivative of the sliding variable, as can be observed by differentiating Eq. \eqref{eq:sigma} and substituting the system dynamics \eqref{eq:dynkinsys}. This ensures that a first-order SMC approach is appropriate for the given problem formulation, as it allows direct control of the sliding dynamics.} Therefore, the control input $u\in\R^3$ is defined according to the first-order SMC algorithm (Eqs. \ref{eq:ctrl_law}-\ref{eq:sat}):}
    u=&-\Gamma \text{sat}(\sigma),\tilda \Gamma=\text{diag}\left(\gamma\mathbf{1}_3\right),\tilda \gamma\in\R^+,
    \label{eq:ctrl_law}\\
    &\text{sat}(\sigma)=\begin{cases}
        \text{sign}(\sigma)\tilda&\text{if}\tilda\left|\sigma\right|\geq S\\
        \frac{\sigma}{S}\tilda&\text{if}\tilda\left|\sigma\right|<S
    \end{cases},\tilda S=\frac{\overline{\sigma}}{\sqrt{3}},\tilda \overline{\sigma}\in\R^+\label{eq:sat},
\end{align}
where $\Gamma\in\R^{3\times3}$ is diagonal positive definite matrix with the vector $\gamma\mathbf{1}_3$ on the diagonal. Also, $sat(\sigma)$ is used in place of $sign(\sigma)$ typically found in the classical SMC law in order to avoid chattering (high-frequency switching near the equilibrium point $\sigma=0$ due to discontinuity of signum function).
This is the boundary-layer SMC approach, which worsens the tracking accuracy as the closed loop system can not converge on $\sigma=0$ as guaranteed by classical SMC approach. Instead, the closed-loop system under boundary-layer SMC approach converges to the boundary layer $\|\sigma\|\leq\overline{\sigma}$, where $\overline{\sigma}>0$ gives the width of the boundary layer. Note that both the functions $sign$ and $sat$ apply to each element of $\sigma$, so that they belong to $\R^3$.

Now, the time evolution of closed-loop system is analyzed by considering two sequential stages by means of relevant Propositions as follows. The first stage  {is addressed by exploiting Lemma \ref{lemma:FTS}} and concerns the reachability of the boundary layer neighbouring the sliding manifold in finite time (reaching phase). 
Then, the second stage deals with evolution of the trajectories within the boundary layer (sliding phase)  and {is addressed by exploiting Lemma \ref{lemma:AS}}. 

\begin{proposition}
    Consider the dynamics system (\ref{eq:dynkinsys}) with the control law $u$ (\ref{eq:ctrl_law}-\ref{eq:sat}) producing the control torque $\tau=I^*u$. Also, consider the sliding variable (\ref{eq:sigma}) with $q^*$ as in Eq. (\ref{eq:q^*}) and $\omega^*$ resulting from Eq. (\ref{eq:omega^*}).
     {Following Proposition 1, consider only one forbidden zone $(j=1)$ contributing to the repulsive factor in $\omega^*$, so that $\|\omega^*\|\leq\overline{\omega}$ holds at any time of the control process. 
    Now, suppose that the states of system (\ref{eq:dynkinsys}) fulfills $\|\omega\|\leq\overline{\omega}$, so that the trajectories evolve along the set
    \begin{equation}
        \mathcal{S} = \left\{ (q, \omega):\tilda q \in \mathbb{R}^4, \; \|\omega\| \leq \overline{\omega} \right\},
        \label{eq:set_sys}
    \end{equation}}
    and consider the assumptions in Section \ref{subsec:Assumption}.
    Further assuming that the actuation system produces a maximum torque value $\overline{\tau}$ satisfying
    \begin{equation}
        \overline{\tau}>\|I^*\|\frac{\left(\|I^{*^{-1}}\|+\|\overline{\Hat{\Delta_{I}}}\|\right)\left(\overline{\omega}\overline{H}+\overline{d}\right)+ {\psi\overline{\omega}^2+\sqrt{2}\lambda\overline{\omega}}}{1- {\|\overline{\Hat{\Delta_{I}}}\tilda I^{*}\|_1}},
    \label{eq:tau_cond}\end{equation}    
    and considering
    \begin{subequations}\begin{align}
        \gamma&=k \frac{\left(\|I^{*^{-1}}\|+\|\overline{\Hat{\Delta_{I}}}\|\right)\left(\overline{\omega}\overline{H}+\overline{d}\right)+ {\psi\overline{\omega}^2+\sqrt{2}\lambda\overline{\omega}}}{1-{\|\overline{\Hat{\Delta_{I}}}\tilda I^{*}\|_1}}\in\R^+,\label{eq:gamma}\\
        k&\in\left(1,\tilda\frac{\overline{\tau}}{\|I^*\|}\frac{1-{\|\overline{\Hat{\Delta_{I}}}\tilda I^{*}\|_1}}{\left(\|I^{*^{-1}}\|+\|\overline{\Hat{\Delta_{I}}}\|\right)\left(\overline{\omega}\overline{H}+\overline{d}\right)+ {\psi\overline{\omega}^2+\sqrt{2}\lambda\overline{\omega}}}\right),\label{eq:k}\\
        \psi&=1+\frac{1+\overline{\epsilon_e}}{2\overline{\epsilon_e}}+\frac{1+\underline{\Tilde{\epsilon}}}{\underline{\Tilde{\epsilon}}}\label{eq:psi}
    \end{align}\label{eq:GammaKappa}\end{subequations}
    the trajectories of the closed-loop system system (\ref{eq:dynkinsys}), (\ref{eq:sigma})-(\ref{eq:sat}), converge in finite time $t_r < \infty$ to the set 
     {\begin{equation}
        \mathcal{S}_{{\sigma}}=\gl(q,\omega):\tilda\|\sigma\|\leq\overline{\sigma}\gr
        \label{eq:set_ov_Sigma}
    \end{equation}}
    and stay there for any $t\geq t_r$. Also, the control input $u$ produced by Eqs. (\ref{eq:ctrl_law}-\ref{eq:sat}) with control gains (\ref{eq:GammaKappa}) satisfies $\|\tau\|<\overline{\tau}$.
\label{prop:RF}

    \begin{proof} \textit{of Proposition \ref{prop:RF}}

    Before giving the details of the proof, note that the set of admissible $k-$values is non-empty thanks to assumption (\ref{eq:tau_cond}). 
     {Furthermore, the assumption $\|\omega\|\leq\overline{\omega}$ is required to preserve the controllability of the satellite, as explained in Section 3.2.}
    Then, a quadratic Lyapunov function {$V_\sigma$} is selected and its time derivatives are computed as follows:
    \begin{equation}
        V_\sigma=\sigma^T\sigma,\quad\Dot{V}_\sigma=2\sigma^T\Dot{\sigma}.
    \label{eq:dotV_sigma}\end{equation}    
    {$V_\sigma$ satisfies Eq. (\ref{eq:V_AS}) of Lemma \ref{lemma:AS}, while } 
   $\Dot{\sigma}$ results from Eq. (\ref{eq:sigma}) as $\Dot{\sigma}=\Dot{\omega}-\Dot{\omega}^*+\Lambda\Dot{e}_\epsilon$. First, $\Dot{\omega}$ is obtained by inserting $\tau=I^*u$ with $u$ as in Eqs. (\ref{eq:ctrl_law}), (\ref{eq:sat}) into Eq. (\ref{eq:dynkinsys}). Also, the Matrix inversion Lemma is used to write as follows:
    \begin{equation}\begin{split}
        \Dot{\omega}&=\left(I^{*^{-1}}+\Hat{\Delta_{I}}\right)\left(-\omega^\times H-I^*\Gamma\text{sat}(\sigma)-\Lambda e_\omega+d\right)=\\
        &=-\left(\mathbb{I}_3-\Hat{\Delta_{I}}I^{*}\right)\Gamma\text{sat}(\sigma)+\left(I^{*^{-1}}+\Hat{\Delta_{I}}\right)\left(-\omega^\times H+d\right).
    \end{split}       
    \label{eq:dot_omega}\end{equation}
    Then, $\Dot{\omega}^*$ in $\mathcal{B}$ frame is obtained as follows from Eq. (\ref{eq:omega^*}):
    \begin{equation}
        \Dot{\omega}^*=-\omega^\times R(q)\left(\omega_a+\omega_{ r }\right)+R(q)\frac{\partial }{\partial t}\left(\omega_a+\omega_{ r }\right),
    \label{eq:dot_omega^*}\end{equation}
    because it is considered only a single forbidden zones impacting $\omega^*$ in accordance with Proposition 1. Finally, $\Dot{e}_\epsilon$ results as follows from the kinematics equations:
    \begin{equation}
        \Dot{e}_\epsilon=\frac{1}{2}\left(e_\eta e_\omega-e_\omega^\times e_\epsilon\right).
    \label{eq:dot_e_eps}\end{equation}
    By inserting Eqs. (\ref{eq:dot_omega})-(\ref{eq:dot_e_eps}) into $\Dot{\sigma}$ above,     Eq. (\ref{eq:dotV_sigma}) yields:
    \begin{equation}\begin{split}
    \Dot{V}_\sigma=&-2\sigma^T\left(\mathbb{I}_3-\Hat{\Delta_{I}}I^{*}\right)\Gamma\text{sat}(\sigma)+\\
    &+2\sigma^T\Big(\left(I^{*^{-1}}+\Hat{\Delta_{I}}\right)\left(-\omega^\times H+d\right)+\\
    &-\omega^\times R(q)\left(\omega_a+\omega_{ r,j }\right)+R(q)\frac{\partial }{\partial t}\left(\omega_a+\omega_{ r,j }\right)+\\
    &+\frac{1}{2}\Lambda\left(e_\eta e_\omega-e_\omega^\times e_\epsilon\right)    \Big).
    \end{split}\label{eq:dot_V_s2}\end{equation}
    Now, note that it exists a known $\overline{\hat{\Delta_I}}>0$ such that $-\overline{\hat{\Delta_I}}<\hat{\Delta_I}<\overline{\hat{\Delta_I}}$ for assumptions made in Section 2.1.1. Furthermore, consider that for $|\sigma|\geq S$ it follows that: (i) $\|\sigma\|\geq\overline{\sigma}$ due to the selection of $S$ of Eq. (\ref{eq:sat}), and (ii) $\sigma^T\cdot sat(\sigma)=\|\sigma\|_1\geq\|\sigma\|$. Therefore, the first line of Eq. (\ref{eq:dot_V_s2}) yields
    \begin{equation}\begin{split}
        &-\|2\sigma^T\left(\mathbb{I}_3-\Hat{\Delta_{I}}\tilda I^{*}\right)\Gamma\text{sat}(\sigma)\|_1\leq-2\gamma\left(1-\|\overline{\Hat{\Delta_{I}}}\tilda I^{*}\|_1\right)\|\sigma\|<0
    \end{split}\label{eq:fot_V_s_1st_row}\end{equation}
    because $\|\Hat{\Delta_{I}}\tilda I^{*}\|_1<1$ due to $\overline{\Delta_I}<I^*$ and $\gamma>0$. 
     {Then, the other terms of $\dot{V}_\sigma$ are manipulated as follows. The second line of Eq. \eqref{eq:dot_V_s2} is bounded as follows:
        \begin{equation}\begin{split}
            &2\sigma^T\left(I^{*^{-1}}+\Hat{\Delta_{I}}\right)\left(-\omega^\times H+d\right)\leq\\
            \leq&2\|\sigma^T\|\tilda\|\left(I^{*^{-1}}+\Hat{\Delta_{I}}\right)\left(-\omega^\times H+d\right)\|\leq\\
            \leq&2\left(\|I^{*^{-1}}\|+\|\overline{\Hat{\Delta_{I}}}\|\right)\left(\overline{\omega}\overline{H}+\overline{d}\right)\|\sigma\|
        \end{split}\end{equation}
        due to validity assumption $\|\omega\|\leq\overline{\omega}$, actuation constraint $(\|H\|\leq\overline{H})$, and assumptions in Section 3.1.1 ($\Hat{\Delta_I}<\overline{\hat{\Delta_I}}$, $\|d\|\leq\overline{d}$). 
        Instead, the third line of Eq. \eqref{eq:dot_V_s2} is manipulated as follows:
        \begin{equation}\begin{split}
            &2\sigma^T\left(-\omega^\times R(q)\left(\omega_a+\omega_{ r,j }\right)+R(q)\frac{\partial }{\partial t}\left(\omega_a+\omega_{ r,j }\right)\right)\leq\\
            \leq&2\|\sigma^T\|\left(\|\omega\|\left(\|\omega_a\|+\|\omega_{ r,j }\|\right)+\|\frac{\partial \omega_a}{\partial t}\|+\|\frac{\partial \omega_{ r,j }}{\partial t}\|\right)
        \end{split}\end{equation}
        due to the orthogonality of the rotation matrices ($\|R(q)\|=1$). Also, the validity assumptions of Proposition \ref{prop:RF} enable $\|\omega\|\leq\overline{\omega}$ and $\|\omega^*\|\leq\overline{\omega}$, so that 
        $$
        \|\omega\|\left(\|\omega_a\|+\|\omega_{ r,j }\|\right)\leq\overline{\omega}^2.
        $$
        Then, from Eq. \eqref{eq:omega_a} it follows that:
        \begin{itemize}
            \item $\|\epsilon_e\|\leq \overline{\epsilon_e}$: 
            \begin{equation}
                \frac{\partial \omega_a}{\partial t}={\alpha_1}\dot{\epsilon_e}\tilda\text{sign}_+(\eta_e)=\frac{\alpha_1}{2}\left(\eta_e\omega-\omega^\times\epsilon_e\right)\tilda\text{sign}_+(\eta_e).
                \label{eq:d_om_a_dt}
            \end{equation}
            Due to $\|\omega\|\leq\overline{\omega}$, $\alpha_1=\alpha_2/\overline{\epsilon_e}$, $\alpha_2=\overline{\omega}$, and $\|\epsilon_e\|\leq \overline{\epsilon_e}$, Eq. \eqref{eq:d_om_a_dt} yields as follows:
            \begin{equation}
                \|\frac{\partial \omega_a}{\partial t}\|=\frac{\alpha_1}{2}\overline{\omega}\left(\left|\eta_e\right|+\|\epsilon_e\|\right)\leq\frac{\overline{\omega}^2}{4}\frac{1+\overline{\epsilon_e}}{\overline{\epsilon_e}}.
            \end{equation}
            \item $\|\epsilon_e\|> \overline{\epsilon_e}$: 
            \begin{equation}\begin{split}
                 &\frac{\partial \omega_a}{\partial t}=\alpha_2\left(\frac{\dot{\epsilon_e}}{\|\epsilon_e\|}-\frac{\epsilon_e\left(\epsilon_e^T\dot{\epsilon_e}\right)}{\|\epsilon_e\|^3}\right)\tilda\text{sign}_+(\eta_e)\tilda\Rightarrow\\
                 \Rightarrow\tilda&\|\frac{\partial \omega_a}{\partial t}\|\leq\alpha_2\left(\frac{\|\dot{\epsilon_e}\|}{\|\epsilon_e\|}-\frac{\|\epsilon_e\|\|\epsilon_e^T\dot{\epsilon_e}\|}{\|\epsilon_e\|^3}\right)\leq2\alpha_2\frac{\|\dot{\epsilon_e}\|}{\|\epsilon_e\|}\leq\\
                 \leq\tilda&\alpha_2\overline{\omega}\frac{|\eta_e|+\|\epsilon_e\|}{\|\epsilon_e\|}\leq\frac{\overline{\omega}^2}{2}\frac{1+\overline{\epsilon_e}}{\overline{\epsilon_e}}.
            \end{split}\end{equation}
        \end{itemize}
        Instead, from Eq. \eqref{eq:Ur} it is obtained as follows:
        \begin{equation}\begin{split}
            &\frac{\partial \omega_{r,j}}{\partial t}=\frac{\partial }{\partial t}\left(\frac{\zeta}{\|\Tilde{\epsilon}_j\|^3}\right)\Tilde{\epsilon}_j+\frac{\zeta}{\|\Tilde{\epsilon}_j\|^3}\frac{\partial \Tilde{\epsilon}_j}{\partial t}=-3\zeta\frac{\Tilde{\epsilon}_j^T\dot{\Tilde{\epsilon}}_j}{\|\Tilde{\epsilon}_j\|^5}\Tilde{\epsilon}_j+\frac{\zeta}{\|\Tilde{\epsilon}_j\|^3}\dot{ \Tilde{\epsilon}}_j\tilda\Rightarrow\\
            \Rightarrow&\|\frac{\partial \omega_{r,j}}{\partial t}\|\leq3\frac{\zeta}{\|\Tilde{\epsilon}_j\|^3}\|\dot{\Tilde{\epsilon}}_j\|+\frac{\zeta}{\|\Tilde{\epsilon}_j\|^3}\|\dot{\Tilde{\epsilon}}_j\|\leq\\
            &\quad\quad\quad\leq2\zeta\|{\omega}\|\frac{|\Tilde{\eta}_j|+\|\Tilde{\epsilon}_j\|}{\|\Tilde{\epsilon}_j\|^3}\leq2\zeta\overline{\omega}\frac{1+\underline{\Tilde{\epsilon}}}{\underline{\Tilde{\epsilon}}^3}=\overline{\omega}^2\frac{1+\underline{\Tilde{\epsilon}}}{\underline{\Tilde{\epsilon}}}
        \end{split}\end{equation}
    Then, the last line of Eq. \eqref{eq:dot_V_s2} yields as follows:
        \begin{equation}\begin{split}
            \sigma^T\Lambda\left(e_\eta e_\omega-e_\omega^\times e_\epsilon\right)\leq&\|\sigma\|^T\|\Lambda\|\|e_\omega\|\left(|e_\eta|+\|e_\epsilon\|\right)\leq\\
            \leq&\|\sigma\|^T\lambda \sqrt{2} \|e_\omega\|\leq 2\sqrt{2} \lambda\overline{\omega}\|\sigma\|
        \end{split}\label{eq:dot_V_s_last_row}\end{equation}
        due to $|e_\eta|+\|e_\epsilon\|=\sqrt{1-\|e_\epsilon\|^2}+\|e_\epsilon\|\leq\sqrt{2}$ and         
        $\|e_\omega\|=\|\omega-\omega^*\|\leq2\overline{\omega}$. By inserting Eqs. (\ref{eq:fot_V_s_1st_row}-\ref{eq:dot_V_s_last_row}) into Eq. \eqref{eq:dot_V_s2}, it is obtained as follows:
            \begin{equation}\begin{split}
    \Dot{V}_\sigma\leq&2\bigg(-\gamma\left(1-\|\overline{\Hat{\Delta_{I}}}\tilda I^{*}\|_1\right)+\left(\|I^{*^{-1}}\|+\|\overline{\Hat{\Delta_{I}}}\|\right)\left(\overline{\omega}\overline{H}+\overline{d}\right)+\\
    &+\overline{\omega}^2+\frac{\overline{\omega}^2}{2}\frac{1+\overline{\epsilon_e}}{\overline{\epsilon_e}}+{\overline{\omega}^2}\frac{1+\underline{\Tilde{\epsilon}}}{\underline{\Tilde{\epsilon}}}+\sqrt{2}\lambda\overline{\omega}    \bigg)\|\sigma\|.
    \end{split}\label{eq:last_RF_proof}\end{equation}
    By inserting $\gamma$ as in Eq. \eqref{eq:GammaKappa}, equation above yields $\Dot{V}_\sigma\leq-2(k-1) V_\sigma^{\nicefrac{1}{2}}$ for any $\|\sigma\|>\overline{\sigma}$, with $2(k-1)>0$ guaranteed by Eq. \eqref{eq:k}. Therefore, Eq. \eqref{eq:last_RF_proof} realizes condition (\ref{eq:dotV_FTS}),  Lemma \ref{lemma:FTS} applies, and the trajectories of the closed-loop system converge in finite-time to the set \eqref{eq:set_ov_Sigma}. }

     Consider now the torque saturation problem, which is avoided if the control torque $\tau$ fulfills $\|\tau\|\leq\overline{\tau}$ at any time of the control process. In Section 3.1, the control torque was defined as $\tau=I^*u$, thus implying $\|\tau\|\leq\|I^*\|\|u\|$. Therefore, the torque saturation is avoided if $\|u\|\leq\overline{\tau}/\|I^*\|$. 
    The control law (\ref{eq:ctrl_law})-(\ref{eq:sat}) produce a control input $u$ such that $\|u\|=\|\Gamma\|=\gamma$. Introducing the upper bound of the admissible values of $k$ (Eq. \ref{eq:k}) into the definition of $\gamma$  (Eq. \ref{eq:gamma}), it can be found that $\gamma$ is upper bounded as $\gamma<\overline{\tau}/\|I^*\|$. The latter realizes $\|u\|<\overline{\tau}/\|I^*\|$ due to $\|u\|=\gamma$, so that 
    $\|\tau\|\leq\overline{\tau}$ is verified as $\|\tau\|<\|I^*\|\tilda\|u\|<\|I^*\|\tilda\overline{\tau}/\|I^*\|=\overline{\tau}$. 
\end{proof}\end{proposition}
Proposition \ref{prop:RF} and relevant proof state that $\exists\tilda t_r$ such that $\|\sigma\|\leq\overline{\sigma}\tilda\forall\tilda t\geq t_r$. Therefore, the trajectories of the system for $t\geq t_r$ (sliding phase) are constrained to evolve according to the following homogeneous linear time-invariant
differential equation resulting from Eq. (\ref{eq:sigma}):
\begin{equation}
    e_\omega+\Lambda e_\epsilon=\sigma,\quad\|\sigma\|\leq\overline{\sigma}
\label{eq:reduced_dyn}\end{equation}
Before addressing the time-evolution of the system sliding phase, let us point out the following Remark:
\begin{remark}
    Since the reference quaternion $q^*$ is obtained by propagating the actual quaternion $q$ with the reference angular speed $\omega^*$, and the last accounts for the unwinding problem as discussed in Section 3.1, the quaternion error $e_q=\overline{q^*}\otimes q$ in Eq. (\ref{eq:errors}) is such that $e_\eta\in\R^+$.
\label{remark}\end{remark}
\begin{proposition}
    Take the system (\ref{eq:dynkinsys}), (\ref{eq:errors})-(\ref{eq:sat})  constrained to evolve on the set \eqref{eq:set_ov_Sigma}, which is verified  $\forall t \geq t_r$. Also, take $\lambda\in\R^+$ as in Eq. (\ref{eq:sigma}) and consider the tracking error $e_\epsilon$ as in Eq. (\ref{eq:errors}). Then, $e_\epsilon$ asymptotically converge to the set
    \begin{equation}
         {\mathcal{S}_\epsilon=}\Bigg \gl e_\epsilon:\tilda\|e_\epsilon\|\leq\frac{\overline{\sigma}}{\lambda} \Bigg \gr
    \label{eq:accuracy}\end{equation}
     {as the system evolves according to the reduced dynamics (\ref{eq:reduced_dyn}), which is verified  $\forall t \geq t_r$.}
\begin{proof} \textit{of Proposition \ref{prop:SF}}

Consider the quadratic Lyapunov function $V_\epsilon$ and its time derivatives: 
\begin{equation}
    V_\epsilon=e_\epsilon^T e_\epsilon,\quad\Dot{V}_\epsilon=e_\epsilon^T\left(e_\eta e_\omega-e_\omega^\times e_\epsilon\right).
\end{equation}
     {$V_\epsilon$ satisfies Eq. (\ref{eq:V_FTS}) of Lemma \ref{lemma:AS}, while } 
    $\Dot{e}_\epsilon$ is obtained from Eq. (\ref{eq:dot_e_eps}). Note that the cross product $e_\omega^\times e_\epsilon$ produce a vector perpendicular to $e_\epsilon$, so that $\Dot{V}_\epsilon=e_\epsilon^T e_\eta e_\omega$.
    Now, insert $e_\omega=\sigma-\Lambda e_\epsilon$ Eq. (\ref{eq:reduced_dyn}) into $\Dot{V}_\epsilon$ to obtain the following:
    \begin{equation}
        \Dot{V}_\epsilon=-e_\eta e_\epsilon^T \Lambda e_\epsilon + e_\eta e_\epsilon^T\sigma.
    \label{dot_V_e_2}\end{equation}
    Note that $e_\eta\in\R^+$ as pointed out by Remark \ref{remark}, thus $e_\eta e_\epsilon^T \Lambda e_\epsilon<0$ for any $e_\epsilon\neq 0$ due to $\Lambda>0$. Also, $\|e_\epsilon^T \Lambda e_\epsilon\|=\|e_\epsilon\|\gamma\|e_\epsilon\|$.   
    Moreover, $\|\sigma\|\leq\overline{\sigma}$ enables to write $e_\epsilon^T\sigma\leq\|e_\epsilon\|\overline{\sigma}$. Therefore, Eq. (\ref{dot_V_e_2}) becomes as follows:
    \begin{equation}
        \Dot{V}\leq -e_\eta\|e_\epsilon\|\left(\lambda \|e_\epsilon\|-\overline{\sigma}\right)\leq0\tilda\forall\tilda\|e_\epsilon\|\geq\frac{\overline{\sigma}}{\lambda},
    \label{eq:V_eps_last}\end{equation}
    which  {shows Eq. \eqref{eq:dotV_FTS} of Lemma \ref{lemma:AS}, thus proving the } 
    veracity of Proposition \ref{prop:SF}. 
    It remains to discuss the case where $e_\eta=0$ and $\|e_\epsilon\|=1$, in which $\Dot{V}_\epsilon=0$ as evident by Eq. (\ref{eq:V_eps_last}). Actually, this is not an attractor for system (\ref{eq:reduced_dyn}), because it would return $\pm e_{\omega_{i}}\in\left[\lambda-\overline{\sigma},\tilda\lambda+\overline{\sigma}\right]$ for each component $i=1,2,3$ of $e_\omega$. Therefore, the selection $\overline{\sigma}<\lambda$ implies $e_\omega\neq0$ for $e_\eta=0$, i.e. $\Dot{e}_\epsilon\neq0$ by virtue of Eq. (\ref{eq:dot_e_eps}), meaning that the trajectories of system (\ref{eq:reduced_dyn}) escape from $e_\eta=0$ and $\|e_\epsilon\|=1$. Also, note that the selection $\overline{\sigma}<\lambda$ is necessary to provide the control system with sensible tracking accuracy in accordance with Eq. (\ref{eq:accuracy}).
\end{proof}\label{prop:SF}\end{proposition}
    In order to conclude this Section, some comments are provided below. First, Proposition \ref{prop:SF} can be proved true even without relying on Remark \ref{remark}, provided the sliding surface is modified as follows: $\sigma_s=e_\omega+sign(e_\eta)\Lambda e_\epsilon$. With this modification, Eq. (\ref{eq:V_eps_last}) is verified even for $e_\eta\in\R^-$. Furthermore, the situation $e_\eta=0$ and $\|e_\epsilon\|=1$ is unlikely to occur since ${q^*}$ is obtained by propagating the actual quaternion $q$ with the reference angular speed $\omega^*$, while the last one is usually small due to actuators limitations.

\begin{figure*}[t]
    \centering
    \includegraphics[width=0.8\textwidth]{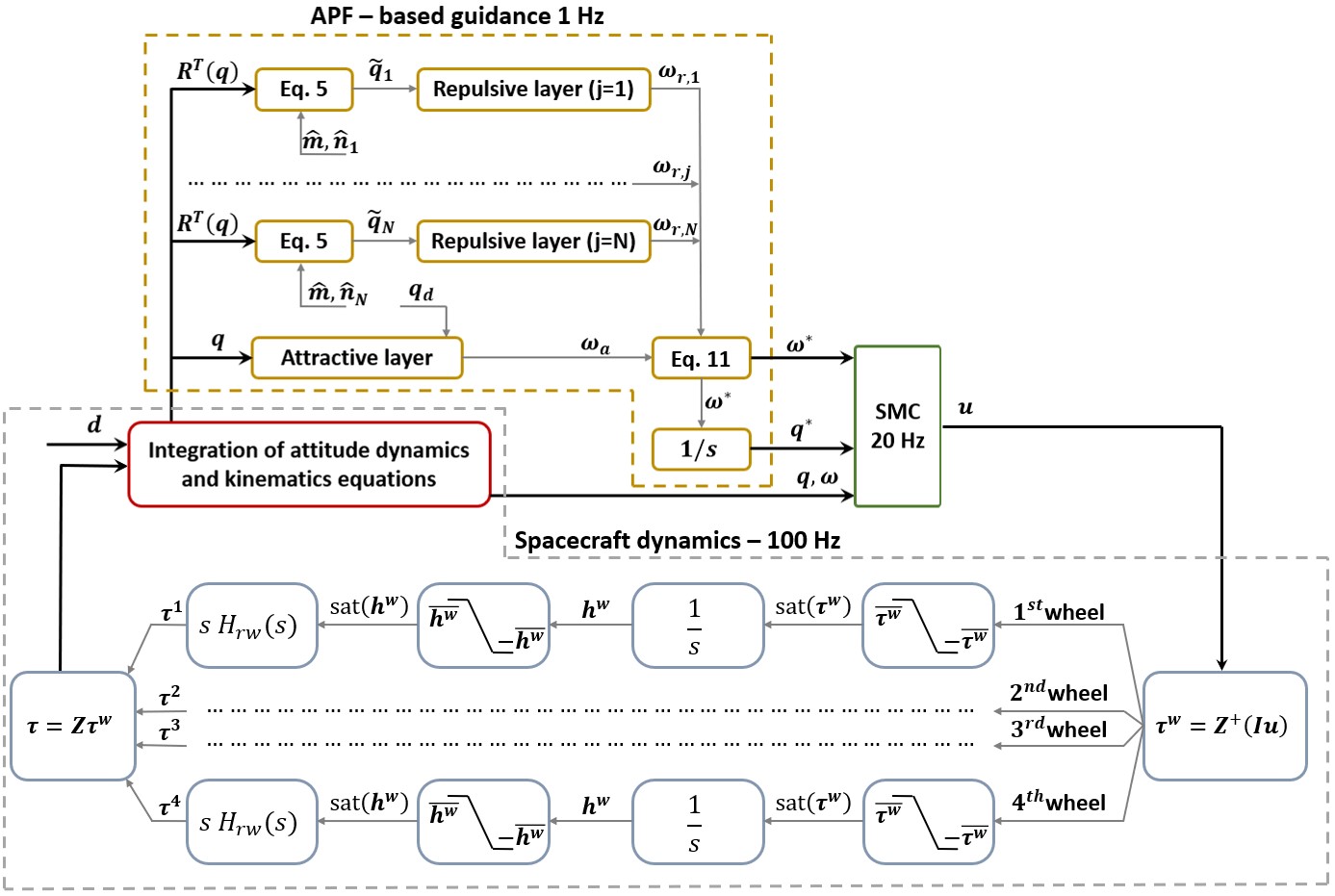}
    \caption{Scheme of the orbital simulator.}
    \label{fig:orbital_simu}
\end{figure*}

\subsection{Robust stability verification}
\label{sc:rob_stab}
\begin{figure}
    \centering
    \includegraphics[width=.4\columnwidth]{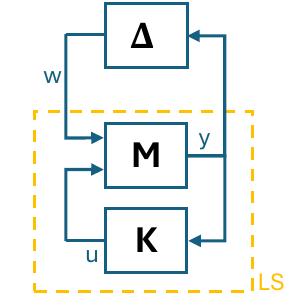}
    \vspace{0ex}
    \caption{Standard $M-\Delta$ configuration}
    \label{fig:M-D model}
    \vspace{0ex}
\end{figure}

Robust stability verification is carried out using $\mu$-analysis. It consists of a typical linear tool for stability verification.  First, the system is exploited by the linear fractional transformation, as in \cite{doyle1982analysis}. As shown in Fig. \ref{fig:M-D model}, the standard configuration $M-\Delta$  is considered, where the uncertainty matrix $\Delta(s)$ is connected to the given system $M(s)$, defining the feedback connection $F(M,\Delta)$. Following the Structured Singular Value (SSV) principle, the feedback connection $F(M,\Delta)$ robust stability is guaranteed if
\begin{equation}
    \sup_{\tilde{\omega}\in\mathbb{R}}\{\mu_\Delta(M(j\tilde{\omega}))\}\leq1
    \label{eq:mu_margin}
\end{equation}
at each given frequency $\tilde{\omega}$, where $\mu_\Delta(M)$ is the SSV, and it is defined as:
\begin{equation}
    \mu_\Delta(M)=\frac{1}{\sup_{\Delta\in\Delta^*}\{\Delta\ |\ \text{det}(I-M\Delta)\neq0\}},
\end{equation}
where $\Delta^*$ is the set of values that takes account of the possible bounded variation of $\Delta$. In practice, it is not always possible to analytically compute the SSV margin expressed in Eq. \eqref{eq:mu_margin}. For this reason, the typical approach is to rely on numerical approaches, as shown in \cite{gu2005robust,young1991mu}.

The robust stability of the closed-loop system is discussed considering $\Dot{\omega}^*=0$ and the system (\ref{eq:dynkinsys}) linearized around $\omega=0,\tilda q=[1,\tilda\mathbf{0}_3^T]$. 
These assumptions are justified by practical considerations. First of all, satellites actuated by RWs typically exhibit sufficiently slow dynamics ($\overline{\omega}$ small). Furthermore, the APF yields constant $\omega^*$ except in the neighborhood of forbidden zones and of desired attitude. Finally, assuming proper tracking behaviour from the beginning of the control process, it is reasonable to consider small quaternion errors, as they are calculated by propagating the initial conditions with $\omega^*$ (small).
Therefore, the following linear system is considered as in \cite{mancini2023adaptive}:
\begin{equation}
    \dot{e}_\epsilon=\frac{1}{2}e_\omega,\quad\dot{e}_\omega=I^{-1}\tau.
    \label{eq:lin_sys}
\end{equation}
Furthermore, Proposition \ref{prop:RF} and relevant proof state that the closed-loop system converges to the boundary layer $\|\sigma\|\leq\overline{\sigma}$. This allows to exploit the SMC control input as a feedback controller, given as
\begin{equation}
    u=-\Gamma\left(\frac{e_\omega+\Lambda e_\epsilon}{\bar{\sigma}}\right).
    \label{eq:lin_contr}
\end{equation}
Note that actual control law interpolates the discontinuous control law for $\sigma_i\in[-S,S]$ ($i=1,2,3$) with $S<\overline{\sigma}$ (Eqs. \ref{eq:ctrl_law}-\ref{eq:sat}), which makes Eq. (\ref{eq:lin_contr}) conservative. 
Linearized closed-loop system is given by combining Eq. \eqref{eq:lin_sys} with Eq. \eqref{eq:lin_contr} through $\tau=I^*u$, where $I^*$ is the nominal value of the inertia tensor. Following this procedure, it is possible to exploit the system in the feedback connection $F(M,\Delta_I)$, where $\Delta_I\in\mathbb{R}^{6\times6}$ is the uncertainty matrix taking into account the uncertainties of the inertia tensor.
Since the nonlinearities in the closed-loop system cannot be neglected in the robustness stability analysis, they are included using the simplified approach proposed in \cite{pagone2020gnc}. In particular, the nonlinear uncertainty is included by evaluating the input-output gain of $\Delta_{NL}$, defined as the induced norm of the nonlinear terms in the closed-loop system.  The input-output relationship with the linearized plant is evaluated through the following equation:
\begin{equation}
    ||\Delta_{NL}||_\infty=\sup_{y\neq0}\frac{||w||_\infty}{||y||_\infty},
    \label{eq:delta_nl}
\end{equation}
which is also described by Fig. \ref{fig:M-D model}. 
This allows to incorporates robust stability analysis with the nonlinearities related to external disturbances and gyroscopic terms (including wheels angular momentum). 
{Rigorous derivation of a global norm bound for $\Delta_{NL}$ is challenging and requires extensive data across an infinite-dimensional space. Indeed, the proposed approach is based on the approximation of the nonlinear terms based on the model physical bounds and the specific characteristics of the guidance and control algorithms. This provides a practical, albeit local, approximation of the nonlinear uncertainties within the expected operational envelope. The nonlinearities are accounted in two ways:
\begin{enumerate}[(I)]
    \item The nonlinearities in system dynamics are incorporated including the gyroscopic term (within the expected operational envelope) and the linear contribution of wheels' angular momentum
    \begin{equation}
        \|\Delta_{nld}\|_\infty=\|I^{*^{-1}}\|_\infty\Bar{h}+2\Bar{\omega} .
        \label{eq:delta_nld}
    \end{equation}
    \item External disturbances are included with control input saturation approximation (using the additive uncertainty representation)
    \begin{equation}
\Bar{\Delta}_{sat}=\|I^{*^{-1}}\|_\infty\left(\frac{\Gamma}{\bar{\sigma}}-\frac{\Bar{\tau}\pm\delta d}{2\Bar{\omega}}\right),
      \label{eq:delta_sat}
    \end{equation}
    with $-\bar{d}\leq\delta d\leq \Bar{d}$.
    This approach provides a conservative representation that encapsulates the expected control behavior in presence of control input saturation.
\end{enumerate}}
{Final formulation of the uncertainty system is given by
\begin{equation}
    {\dot{e}=\begin{bmatrix}
        \dot{e}_\epsilon\\\dot{e}_\omega
    \end{bmatrix}=\begin{bmatrix}
        \frac{1}{2}e_\omega\\
        I^{-1}I^*u
    \end{bmatrix}+(\Delta_{nld}+\Delta_{sat})e}.
    \label{eq:sys_unc}
\end{equation}
where $\Delta_{nld}=\delta_{nld}\mathbb{I}_6$ and $\Delta_{sat}=\delta_{sat}\mathbb{I}_6$ are the diagonal uncertainty matrices, with $-\|\Delta_{nld}\|_\infty\leq\delta_{nld}\leq\|\Delta_{nld}\|_\infty$ and $0\leq\delta_{sat}\leq\Bar{\Delta}_{sat}$. 
This practical approach allows to assess the system’s robustness limits, effectively defining a performance index that quantifies how much uncertainty the closed-loop system can tolerate before becoming unstable.}
However, even if the proposed approach allows to extend the robust stability considerations to the nonlinear system, it is not a rigorous analysis. Therefore, a simulation campaign is carried out to complete the overall verification process. 

\section{NUMERICAL EXAMPLE}
This section describes the simulation scenario and provides the results of the numerical examples.
\subsection{Simulation scenario}
\label{sc:sim_scenario}
All tests are done on an orbital simulator that was implemented in Matlab/Simulink environment and is schematized in Fig. \ref{fig:orbital_simu}.
It simulates the non-linear dynamics of the rigid satellite (Eq. \ref{eq:dynkinsys}), the APF-based guidance (Eq. \ref{eq:omega_a}-\ref{eq:omega^*}), the SMC algorithm (Eqs. \ref{eq:errors}-\ref{eq:sat}), and the RWs cluster (Eq. \ref{eq:wheel2body}) with saturation blocks and transfer function as in \cite{Hrws}:
$$
H_{rw}(s)=\frac{1.214s + 0.7625}{s^2 + 2.40s + 0.7625}
$$
These components are linked as shown in Fig. \ref{fig:orbital_simu} and different switching frequencies are considered to be compliant with the on-board hardware constraints: the update frequency for the APF algorithm is 1 Hz, while for the SMC algorithm is 20 Hz. In simulation environment, the control input (\ref{eq:ctrl_law}) is multiplied by $I^*$ to obtain the three axis control torque $\tau$. Then, the latter is allocated  to the four wheels by using the Moore-Pensore pseudo-inverse $Z^+$ as follows: $\tau^w=Z^+\tau$. The latter is sent as input to the RWs model, which updates the angular momentum by numerical integration of $\Dot{h}^w=-\tau^w$. 
Then, ${h}^w$ and $\tau^w$ are transformed in $\mathcal{B}$ coordinates through Eq. (\ref{eq:wheel2body}) and injected to the plant. 

The orbital disturbances enter into the plant and are built up from a bias $(10^{-6}~Nm)$ and a sinusoidal part with amplitude $5\cdot10^{-5}~Nm$ and radian frequency $10^{-3}~rad/s$, so that $\overline{d}=8.7\cdot10^{-5}~Nm$. The DEMETER CNES microsatellite \cite{PITTET2006661} is considered for both the main body of the spacecraft and the RWs, with $I^*=[30~-3~0;-3~30~-2;0~-2~40]$, $\overline{\tau^w}=5\cdot10^{-3}~Nm$ and $\overline{h^w}=0.12~Nms$. Also, a maximum uncertainty of $\overline{\Delta_I}=20\%\tilda I^*$ is considered for the inertia tensor. The RWs cluster is mounted with $\alpha=45^\circ$ and $\beta=35^\circ$, resulting in $\overline{\tau}=8.2\cdot10^{-3}~Nm$, $\overline{h}=0.1968~Nms$ and $\overline{\omega}=3.7\cdot10^{-3}~rad/s$. Then, three forbidden zones are modeled with $\Hat{n}_1=[-0.497,\tilda 0.713,\tilda -0.495]^T$, $\Hat{n}_2=[0.033,\tilda 0.984,\tilda -0.177]^T$, $\Hat{n}_3=[-0.116,\tilda 0.843,\tilda 0.528]^T$, and $\underline{\theta}=15^\circ$ each. The gains of the APF algorithm are derived as described throughout the article, while $k=1.02$, $\lambda=0.01$ are selected to compute the SMC gain $\gamma=2\cdot10^{-4}$. Instead, the width of the boundary layer is set at $\overline{\sigma}=5\cdot10^{-4}$. 
The initial attitude of the spacecraft is $q_0=[-0.306,~0.530,~0.660,~-0.436]^T,~\omega_0=\mathbf{0}_3$, the wheels are initialized with zero angular momentum, and the boresight vector is taken as $\Hat{m}= 0.5774\cdot\mathbf{1}_3$. The reference direction is assumed to be equal to $\hat{m}$, so that the desired orientation is $q_d=[1,\tilda \mathbf{0}_3^T]$. 
\begin{figure}
    \centering
    \includegraphics[width=.8\columnwidth]{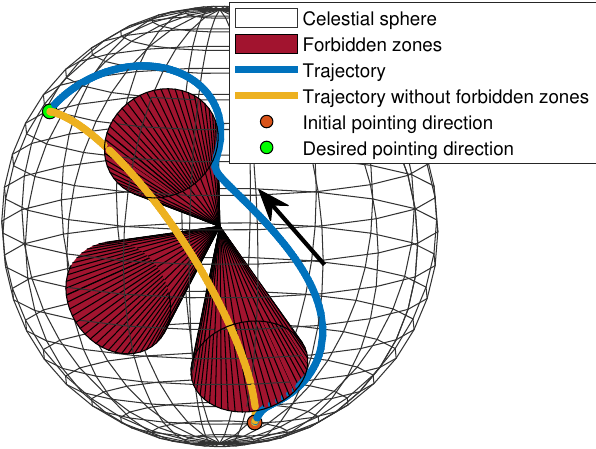}
    \vspace{0ex}
    \caption{Trajectories of boresight vector on the celestial sphere.}
    \label{fig:trajectory}
    \vspace{0ex}
\end{figure}

\subsection{Simulation results}
\begin{figure}
    \centering
    \includegraphics[width=\columnwidth]{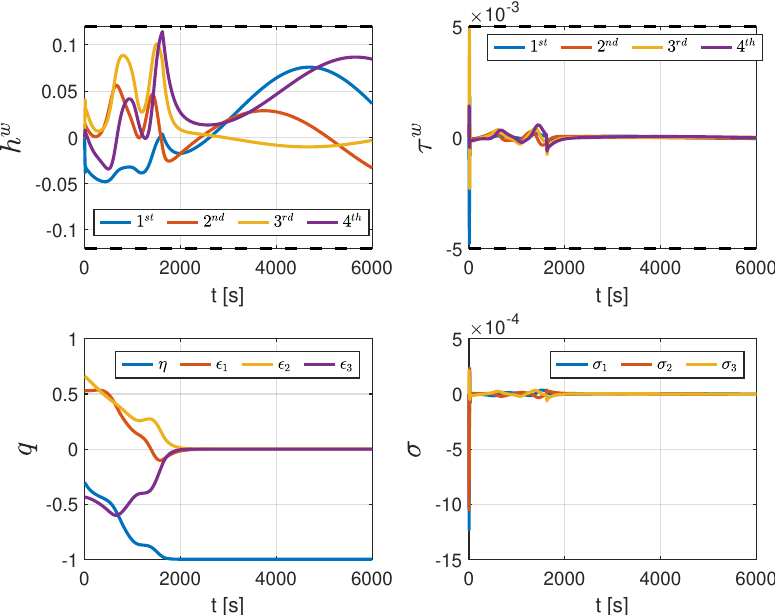}
    \vspace{0ex}
    \caption{Top left and top right: angular momentum and torque produced by each wheel of the cluster, respectively. Bottom left: satellite quaternions evolution. Bottom right: sliding variable.}
    \label{fig:results}
    \vspace{0ex}
\end{figure}

The objective of the simulations is to test the capabilities of the proposed guidance 
and control strategy to perform attitude manoeuvres while dealing appropriately with pointing constraints, external disturbances and actuation constraints. 
Compliance with pointing constraints is analysed through the 3-D slew trajectories of the boresight vector on the celestial sphere (Fig. \ref{fig:trajectory}), the ability to cope with external disturbances is evaluated on the boundedness of the final quaternion error, and fulfilment of the actuation constraints is verified through the time trends of the torque and angular momentum produced by each wheel in the cluster (Fig. \ref{fig:results}).

\begin{figure}
    \centering
    \includegraphics[width=\columnwidth]{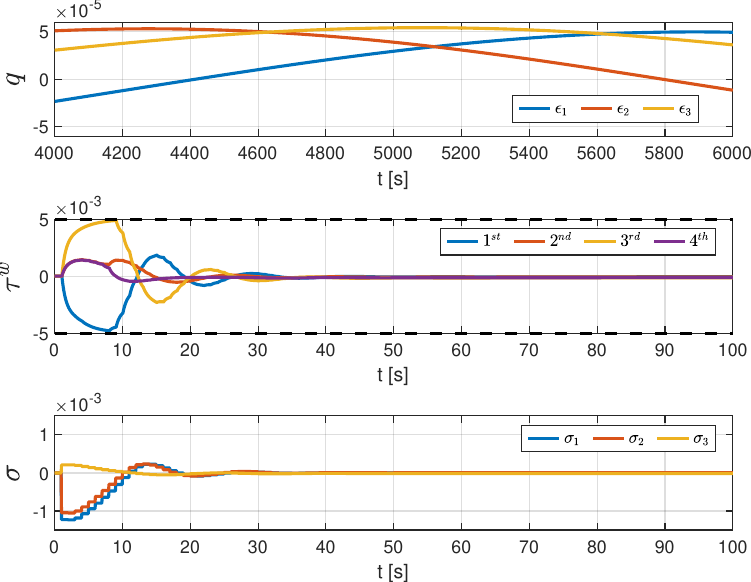}
    \vspace{0ex}
    \caption{ {Top: steady-state error of the vectorial part of the quaternions. Middle: torques produced by each wheel of the cluster during the first 100 s. Bottom: sliding variable evolution for the first 100 s.}}
    \label{fig:results_zoom}
    \vspace{0ex}
\end{figure}

Fig. \ref{fig:trajectory} shows two rotation trajectories linking the initial orientation (red marker) and the desired one (green marker), one depicted in yellow and the other in blue. The yellow trajectory is obtained without considering the forbidden zones in the closed-loop system, so that $\omega^*$ is only produced by the attractive layer of the APF. This case study is of interest to confirm that the attractive potential realises the shorter-path eigenaxis rotation, hence the shortest angular path is covered to reorient the satellite as the yellow line demonstrates. Instead, the blue trajectory is obtained with the forbidden zones in the loop. In this case, the proposed APF/SMC strategy realises a smooth and safe slew trajectory as the boresight vector successfully avoids the forbidden zones before converging on the reference direction.   
Then, the two graphs at the top of Fig. \ref{fig:results} show the torques $\tau^w$ (left, in $Nm$) and angular momentum $h^w$ (right, in $Nms$) developed by each wheel of the cluster. The graphs refer to the maneuver marked by the blue line in Fig. \ref{fig:trajectory}, and it can be seen that the proposed strategy successfully avoids saturations for the entire control process.  {Additionally, a zoom-in of the control torque evolution during the first 100 seconds of the simulation is provided in Fig. \ref{fig:results_zoom}.}  
The bottom left graph of Fig. \ref{fig:results} shows the evolution of the quaternions $q$ , which successfully converges to the identity quaternion during the maneuver as previously evident from Fig. \ref{fig:trajectory}. Moreover, the graph shows that the unwinding problem is successfully solved, as for $\eta_0<0$ the trajectories converge to $\eta=-1$.  
Also, the tracking accuracy in terms of steady-state quaternion error is evaluated in Fig. \ref{fig:results_zoom}, which shows that all components of the vectorial part of the quaternions are confined within a range of $\pm5\cdot10^{-5}$ with respect to the reference, i.e. an error of approximately $10^{-3}$ degrees.
Finally, the bottom right graph of Fig. \ref{fig:results} shows the time evolution of the sliding output $\sigma$ in Eq. (\ref{eq:sigma}), revealing that $\sigma$ is confined throughout the control process. This certifies the effectiveness of the proposed SMC algorithm to steer the satellite along the reference computed by the APF, thus realising a safe and accurate maneuver.  {Additionally, a zoom-in of the sliding variable evolution during the first 100 seconds of the simulation is provided in Fig. \ref{fig:results_zoom}.}

\subsection{$\mu$-analysis}
\begin{figure}
    \centering
    \includegraphics[width=.9\columnwidth]{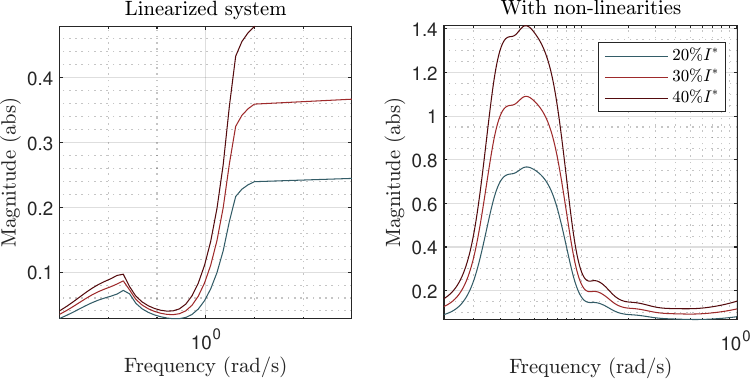}\\
    \includegraphics[width=.9\columnwidth]{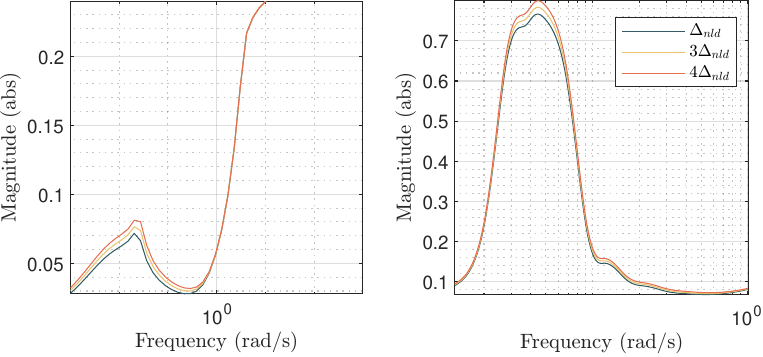}\\
    \includegraphics[width=.9\columnwidth]{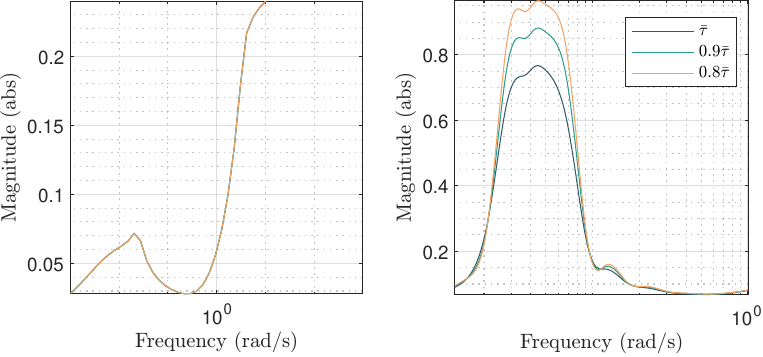}\\
    \includegraphics[width=.9\columnwidth]{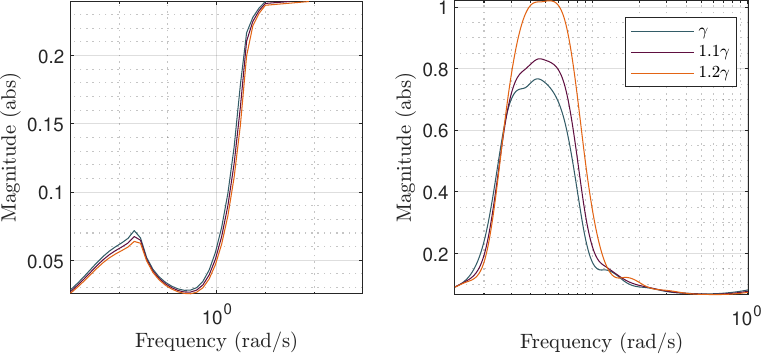}
    \vspace{0ex}
    \caption{$\mu$-analysis: SSV bounds for linerized (left) and nonlinear system (right) for different values of uncerteinties. Blue line is referred to the uncerteinties bounds considered in the design process.}
    \label{fig:MU_analysis}
    \vspace{0ex}
\end{figure}

\begin{figure}
    \centering
    \includegraphics[width=.8\columnwidth]{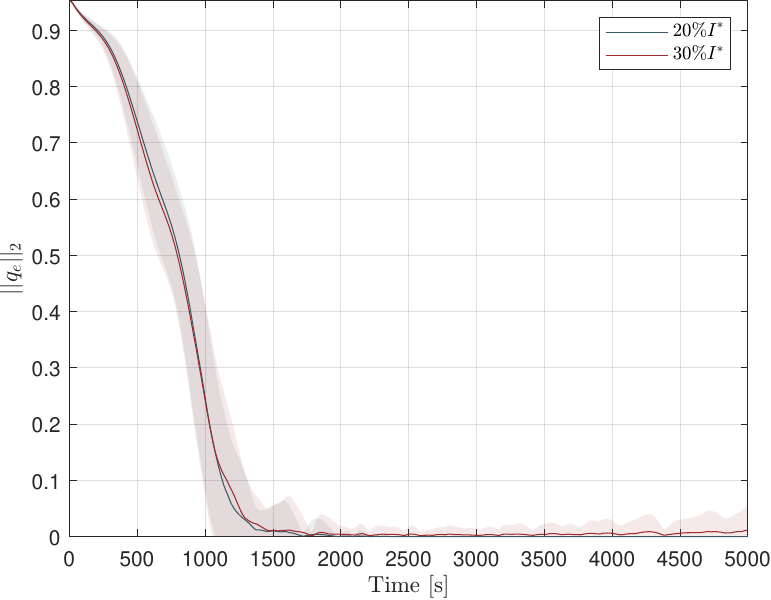}
    \vspace{0ex}
    \caption{Simulation campaign (quaternions error). Numerical verification of closed-loop system sensitivity to inertia matrix uncertainties.}
    \label{fig:MC_MUanalysis}
    \vspace{0ex}
\end{figure}

Robust stability verification {for the uncertain system \eqref{eq:sys_unc}} is carried out through the $\mu$-analyis discussed in Section \ref{sc:rob_stab}{, and  by means of Matlab \textit{mussv} function}. Robust stability is ensured when SSV remains strictly below unity. The resulting upper and lower bounds of the SSV are illustrated in Fig.~\ref{fig:MU_analysis}. The analysis is conducted in two steps. First, a linear fractional transformation of the linearized system is considered, accounting only for uncertainties in the inertia tensor, modeled as $\Delta_I = \pm20\% I^*$. Subsequently, the analysis is extended to the nonlinear system by incorporating uncertainties arising from nonlinear dynamics, denoted as $\Delta_{NL}$. These are computed based on Eqs.~\eqref{eq:delta_nl}–\eqref{eq:delta_sat}, yielding $\|\Delta_{NL}\|_\infty = \|\Delta{nld}\|_\infty + \|\Delta{sat}\|_\infty = 0.5816$. SSV behavior is shown in Fig. \ref{fig:MU_analysis}. The results show that, while high-frequency behavior of the nonlinear system aligns with the linearized model, a significant peak appears at low frequencies, suggesting potential sensitivity issues in steady-state conditions.
{The robustness assessment is performed by evaluating the system response over a range of uncertainty levels, comparing these results with those expected under nominal operating conditions. This approach allows us to determine the maximum level of uncertainty the system can withstand before instability occurs.}
In particular, the blue line (referred to the uncertainties bounds considered in the design process) is considered as reference for closed-loop local sensitivity behavior, while the implication of having different uncertainties is evaluated. The $\mu$-analysis enables the assessment of local stability robustness under the assumed uncertainty bounds and helps identify critical levels of uncertainty. In particular, the critical range for the inertia matrix uncertainty is identified between $20\% I^* < \Delta I^* < 30\% I^*$, beyond which robust stability can no longer be guaranteed. Additionally, the system maintains local stability even under nonlinear uncertainties exceeding three times the expected level, and under a 20\% reduction in available control torque.
Moreover, the effect of different design of $\gamma$ is also evaluated in the lower plot. In particular, two major practical implications can be derived: (i) higher uncertainties makes the system less robust, proving the reliability of the design relationship of $\gamma$, and (ii) increasing the control gain $\gamma$ the system is more sensitive in accordance with the system limitations. It is important to note that the same behavior cannot be observed on the linearized system.
However, the conducted analysis is valid only for structured uncertainties and certain assumed nonlinearities. For this reason, the analysis does not provide any quantitative global conclusion, and simulation campaign is required as part of the verification process.
{In Fig. \ref{fig:MC_MUanalysis}, the performance of the proposed design is evaluated considering a simulation campaign under varying levels of inertia matrix uncertainty, chosen as a representative critical case. Performance is assessed based on the quaternion error, and the results align with the predictions from $\mu$-analysis. When the design assumptions are met, the steady-state error converges is ensured. However, for inertia uncertainties around $30\%I^*$, the system exhibits larger steady-state error, indicating a degraded stability margin. This confirms the robustness limits identified through $\mu$-analysis.}
Deeper insight on the simulation campaign for nominal operating parameters is presented below to complete the verification process.


\subsection{Simulation Campaign}
\begin{figure}
    \centering
    \includegraphics[width=.8\columnwidth]{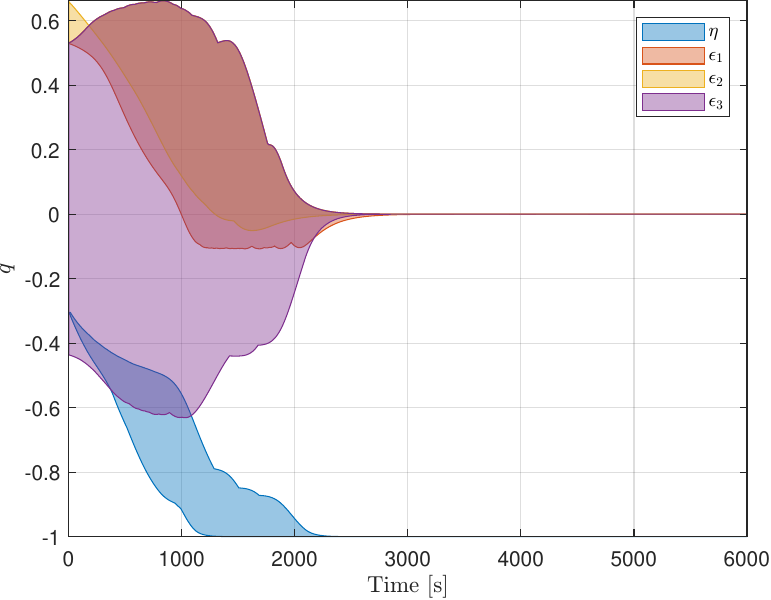}\\
    \includegraphics[width=.8\columnwidth]{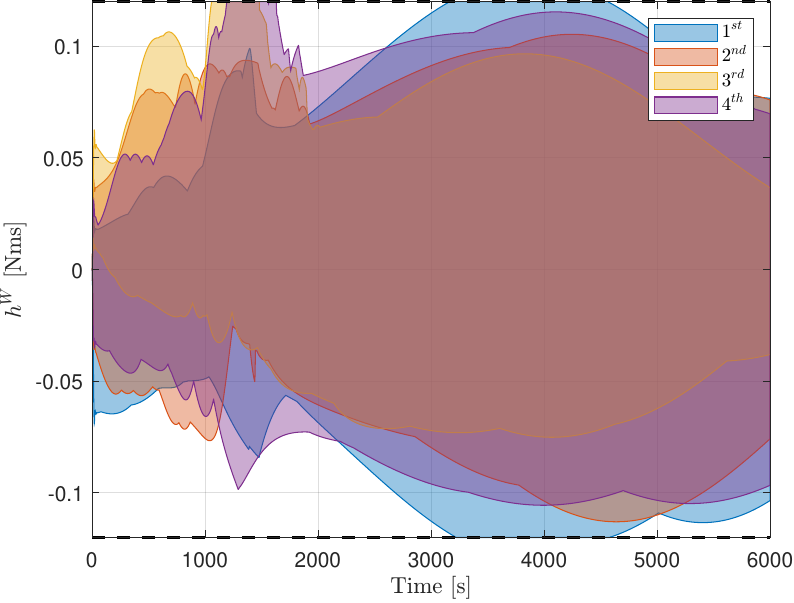}\\
    \includegraphics[width=.8\columnwidth]{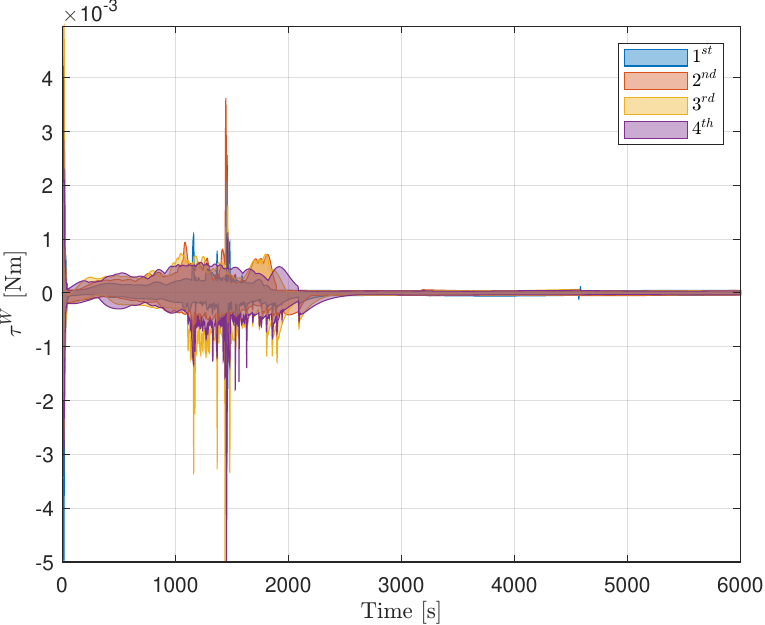}
    \vspace{0ex}
    \caption{Simulation campaign.  Quaternions time variation, on top. Angular momentum and torque produced by each wheel of the cluster, center and bottom, respectively.}
    \label{fig:MC_results1}
    \vspace{0ex}
\end{figure}

\begin{figure}
    \centering
    \includegraphics[width=.8\columnwidth]{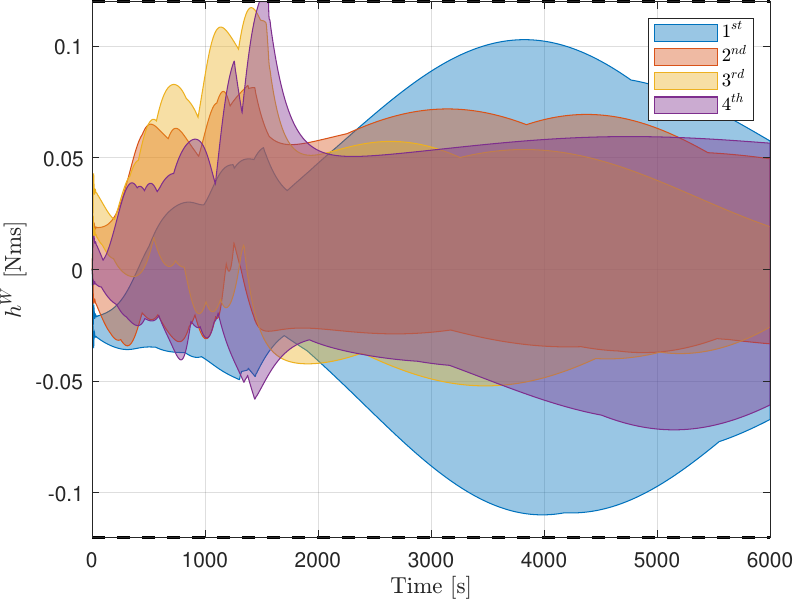}
    \vspace{0ex}
    \caption{Simulation campaign. Reaction wheels' angular momentum in rest-to-rest maneuvers.}
    \label{fig:MC_resultsR2R}
    \vspace{0ex}
\end{figure}

\begin{figure}
    \centering
    \includegraphics[width=.8\columnwidth]{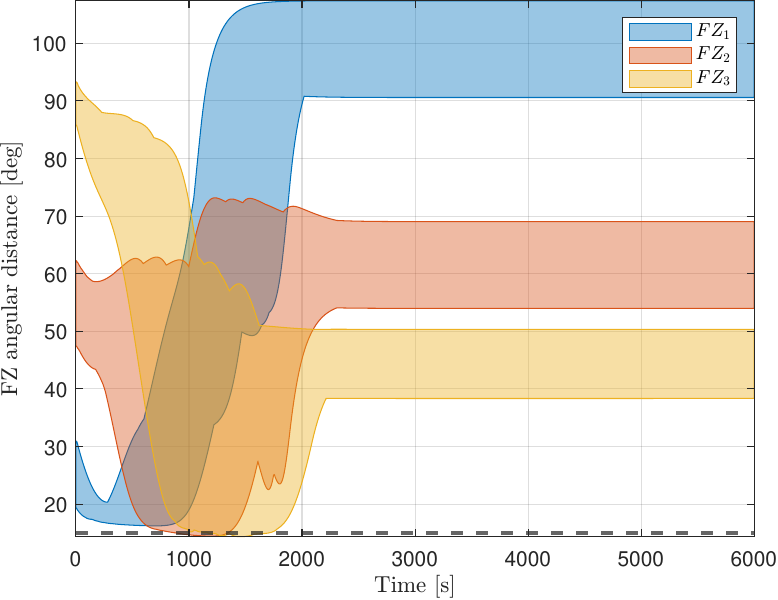}
    \vspace{0ex}
    \caption{Simulation campaign. Minimum angular distance from the forbidden zones.}
    \label{fig:MC_results2}
    \vspace{0ex}
\end{figure}

The simulation campaign is carried out to complete the robust stability verification process, thus evaluating algorithm effectiveness in compliance with system uncertainties (i.e., inertia tensor, external disturbances).
The simulation scenario is designed starting from the conditions given in Section \ref{sc:sim_scenario} (nominal values), while considering randomized variation of the simulator parameters around the nominal values. Parameters value is summarized in Table \ref{tab:sim_param}. In particular, \textit{Simulation parameters} affect the simulation scenario by modifying both initial conditions and location of Forbidden Zones, while \textit{System uncertainties} affect closed-loop system robustness. The simulation campaign consists of 500 simulations with random parameters combination, and it allows to verify empirically closed-loop system robustness. {Initial non-static conditions (with $\omega_0$ satisfying closed-loop stability assumptions) are also included to verify the extendability of the proposed approach to multiple-pointing maneuvers.}
Fig. \ref{fig:MC_results1} shows the variation of quaternions and wheels' torque and angular momentum within the simulation campaign. The results are represented using areas bounding the variation of the considered quantities. The results verify the capability of the proposed approach to successfully accomplish the maneuver, while satisfying wheels saturation requirements. {In particular, this is true for rest-to-rest maneuvers (Fig. \ref{fig:MC_resultsR2R}),} even when the wheels are saturated (for short time) during the control process, it can be noticed that the controller successfully acts avoiding and limiting the duration of the saturation. {For non-static initial conditions, wheels need to cumulate higher values of angular momentum to complete the maneuver.} The simulation campaign allows also to verify the proposed approach successfully addresses forbidden zones avoidance. The minimum angular distance (expressed in deg) between the instrument and the forbidden zones is shown in Fig. \ref{fig:MC_results2}. {Some critical cases are reported (6\%) in which the FZ are violated, with a maximum violation of $0.5\ $deg. Given the operating conditions related to the non-static initial conditions and the overall performance of the control approach, this slight infringement remains within acceptable limits for the application.}
\begin{table}
    \centering
    \caption{Simulation campaign parameters}
    \begin{tabular}{l|c}
    \hline
       Simulation parameters & Range\\
       \hline
       Initial quaternions $\Delta q_0$  & $\pm10\%q_0$ \\
       Initial angular velocity $\Delta\omega_0$ & {$\pm10^{-3}$} rad/s\\
       FZ Direction $\Delta(\hat{\eta}_1,\hat{\eta}_2,\hat{\eta}_3)$ & $\pm15$ deg\\
       \hline
       System uncertainties & Range\\
       \hline
       Inertia tensor $\Delta_I$ & $\pm20\%I^*$\\
       Disturbances d & $\pm\bar{d}$\\
       \hline
    \end{tabular}
    \label{tab:sim_param}
\end{table}

\section{CONCLUSIONS}
This  {paper} proposes an approach based on APF and SMC for spacecraft reorientation with forbidden pointing directions and actuator limitations. The APF/SMC design strategy allows to realize safe attitude maneuvers, fulfilling pointing constraints while avoiding actuators saturations. Indeed, both Lyapunov stability analysis and numerical simulations proved the validity of the proposed guidance and control strategy to manage parameters uncertainties, external disturbances, bounded control, and system non linearities.  
The verification of the closed-loop system robust stability is also {carried out by means of $\mu$-analysis, determining the maximum allowable uncertainty that leads the closed-loop system to instability.} A simulation campaign is carried out as part of the robust verification process, validating empirically $\mu$-analysis results, and evaluating the effectiveness of the method in compliance with system uncertainties. {Moreover, non static initial conditions are considered in the simulation campaign to verify the proposed method to be extendable to multiple-pointing maneuver. However, to ensure practical applicability, cumulative angular momentum dumping should be considered in the pointing sequence planning.}
Future works aim to extend the proposed strategy to manage challenging dynamic pointing constraints, {including reaction wheels desaturation strategy}. Finally, future developments may consider a discrete-time SMC formulation, which can offer improved performance in digital control environments and may be particularly beneficial when dealing with low sampling frequencies. 

 \bibliographystyle{elsarticle-num} 
 \bibliography{biblio}






\end{document}